%% file: main.tex
\documentclass{article} 
\usepackage[final]{colm2026_conference}
\usepackage{microtype}
\usepackage{hyperref}
\usepackage{url}
\usepackage{booktabs}

\usepackage{fontawesome5}

\usepackage{lineno}
\usepackage{url}
\usepackage[T1]{fontenc}
\usepackage{inconsolata}

\usepackage{amsmath, colortbl, booktabs, multirow}
\usepackage{amsmath}
\usepackage{amssymb}
\usepackage{mathtools}
\usepackage{amsthm}

\usepackage{graphicx}
\usepackage{svg}
\usepackage{makecell}

\usepackage{tabularx}
\usepackage[table,xcdraw,dvipsnames,svgnames]{xcolor}
\usepackage{amsmath} 
\usepackage{multirow}
\usepackage{pifont}
\usepackage{booktabs} 
\usepackage[most]{tcolorbox}

\newcommand{\sysname}{{{REVERE}}}

\usepackage{algorithm}
\usepackage{algpseudocode}
\usepackage{amsmath}

\algnewcommand\algorithmicdefine{\textbf{Define:}}
\algnewcommand\DEFINE{\item[\algorithmicdefine]}

\algnewcommand\algorithmicactionspace{\textbf{Action Space:}}
\algnewcommand\ASpace{\item[\algorithmicactionspace]}
\newcommand{\commentq}[1]{%
  \textcolor[HTML]{4F4F4F}{\small$\triangleright$ #1}
}

\newcommand{\grey}[1]{%
  \textcolor[HTML]{4F4F4F}{\small #1}
}

\newcommand{\icheckmark}{\ding{51}}%
\newcommand{\xmark}{\ding{55}}%

\usepackage{tcolorbox}
\tcbuselibrary{listings}
\usepackage[most]{tcolorbox}
\usepackage{float}        
\usepackage{caption}
\usepackage{subcaption}   
\tcbuselibrary{listings}

\usepackage[normalem]{ulem}

\newtcblisting{FullPromptBox}[2]{%
  colback=white,
  colframe=#1,
  title=\textbf{#2},
  listing only,
  breakable,
  listing options={
    breaklines=true,
    basicstyle=\ttfamily\small,
    columns=fullflexible,
    showstringspaces=false,
  }
}

\newtcblisting{NewPromptBox}[2]{%
  colback=white,
  colframe=#1,
  title=\textbf{#2},
  listing only,
  breakable,
  listing options={
    breaklines=true,
    breakautoindent=false,
    breakindent=0pt,
    basicstyle=\ttfamily\small,
    columns=fullflexible,
    showstringspaces=false,
  }
}

\definecolor{modernGreen}{HTML}{63ad77}   
\definecolor{modernRed}{HTML}{ad6365}     
\definecolor{modernGray}{HTML}{6B7280}    

\DeclareRobustCommand{\plus}[1]{%
\ensuremath{{\textcolor{modernGreen}{\scriptstyle +#1}}}%
}

\DeclareRobustCommand{\minus}[1]{%
\ensuremath{{\textcolor{modernRed}{\scriptstyle -#1}}}%
}

\DeclareRobustCommand{\minusg}[1]{%
\ensuremath{{\textcolor{modernGray}{\scriptstyle -#1}}}%
}

\newcommand{\cvpos}[3]{%
\ensuremath{#1_{\textcolor{modernGreen}{\scriptstyle +#2}}}%
}

\newcommand{\cvneg}[3]{%
\ensuremath{#1_{\textcolor{modernRed}{\scriptstyle -#2}}}%
}

\newcommand{\cvposv}[3]{%
\ensuremath{#1_{\textcolor{modernGray}{\scriptstyle \pm #3}}}%
}

\newcommand{\cvnegv}[3]{%
\ensuremath{#1_{\textcolor{modernGray}{\scriptstyle \pm #3}}}%
}

\definecolor{darkblue}{rgb}{0, 0, 0.5}
\hypersetup{colorlinks=true, citecolor=darkblue, linkcolor=darkblue, urlcolor=darkblue}

\title{\sysname: Reflective Evolving Research Engineer}

\author{
\textbf{Balaji Dinesh Gangireddi$^\dagger$}\hspace{2em}
\textbf{Aniketh Garikaparthi$^\dagger$}\\
\hspace{0.3em}\textbf{Manasi Patwardhan$^\dagger$}\hspace{2em}
\textbf{Arman Cohan$^{*}$}
\\[1em]\hspace{0.05em}
$^\dagger$TCS Research \hspace{2em}
$^*$Yale University
\\[0.6em]\hspace{0.05em}
\small{\texttt{dinesh.gangireddi@tcs.com}}
}

\begin{document}

\maketitle

\begin{abstract}
Existing prompt-optimization techniques rely on local signals, causing poor generalization across tasks. In addition, they also rely on weak update mechanisms, such as full-prompt rewrites or unstructured merges, which cause knowledge loss and unstable adaptation. These limitations are magnified in research-coding workflows, which involve heterogeneous repositories and weak feedback, limiting abstraction and learning across tasks. We introduce Reflective Evolving Research Engineer (REVERE), a lightweight, self-adapting agent framework that learns from a Global Training Context, distills recurring cross-repository failure modes into reusable heuristics, and applies targeted, code-based edits to agent prompts. REVERE\footnote{Code : \faGithub\ \url{https://revere-acceleron.github.io/revere/}} is evaluated across settings ranging from long-horizon to single-shot benchmarks, and improves over prior expert-crafted instructions by $4.50\%$ on SUPER, $1.3\%$ on ResearchCodeBench, and $4.89\%$ on ScienceAgentBench. It does so at nearly $10$× lower cost and $2.7$× faster adaptation than existing prompt-optimization baselines, demonstrating that self-adapting agents with continual learning and global memory consolidation can meaningfully evolve their capabilities over time.

\end{abstract}

\input{paper}

\bibliography{references}
\bibliographystyle{colm2026_conference}

\newpage
\appendix

\input{appendix}

\end{document}

%% file: paper.tex
\section{Introduction}

While recent progress of Large language models (LLMs) on short-horizon well-specified coding tasks is promising \citep{swe-agent,livebench,gauthier2024aiderpolyglot}, reliability degrades substantially in research-code reproduction \citep{paperbench,scireplicatebench,researchgym,super,rcb}, due to fundamentally different demands on agents. These include coordinating long-horizon tasks under weak and delayed feedback, inferring tacit assumptions, and accumulating procedural knowledge across heterogeneous research frameworks \citep{whyaiscientistsfails,tacit,wang2026firebenchevaluatingagentsrediscovery}. Prior agentic systems \citep{paperbench, open-hands} targeting research reproducibility, typically rely on static prompts. More complex systems such as \citep{paper2code,lin2025autop2cllmbasedagentframework} further decompose high-level tasks through multi-agent workflows; while this can improve reliability, they still operate within fixed contexts and predefined strategies. As a result, these systems struggle to adapt to the evolving conventions and diverse open-ended nature of research-coding tasks.

Recent works on self-refinement \citep{reflexion, selfrefine, majumder2024clin} improve reasoning through iterative feedback but remain instance-specific, motivating prompt-level and experience-based adaptation methods \citep{gepa, mipro, zhao2024expelllmagentsexperiential}. Yet these adaptation methods rely primarily on heuristic prompt sampling and local evaluation signals, which work well in short-horizon settings but cause overfitting to recent outcomes.
This local-signal problem persists even in recent methods such as \citet{dynamiccheatsheet, ace}. Both operate over a limited window of past experiences, without a holistic view across the experiences accumulated from all prior tasks; ACE \citep{ace}, in particular, retains only a single task's experience at a time. In long-horizon settings, this is especially limiting: a bounded window can't hold enough past experience to tell which strategies actually generalize. But keeping every experience without consolidating it into context isn't practical either, and as a result, these methods can still converge to local optima \citep{localoptima} rather than build genuinely generalizable strategies. A second, distinct issue lies in the edit mechanisms of these frameworks themselves: they fall into two failure modes. Full-prompt regeneration \citep{gepa,dynamiccheatsheet} risks semantic drift and knowledge loss as prompts are rewritten over and over, while append-only accumulation \citep{ace} avoids drift but grows unbounded and risks conflicting heuristics without reconsideration. Structured editing methods \citep{sammo} partially address this tension but typically require more complex implementations. 
This motivates an agent that learns from its own trajectories over time, applying targeted, non-destructive updates to learnings rather than full rewrites.

\begin{figure*}[!t]
    \centering
\includegraphics[width=\textwidth]{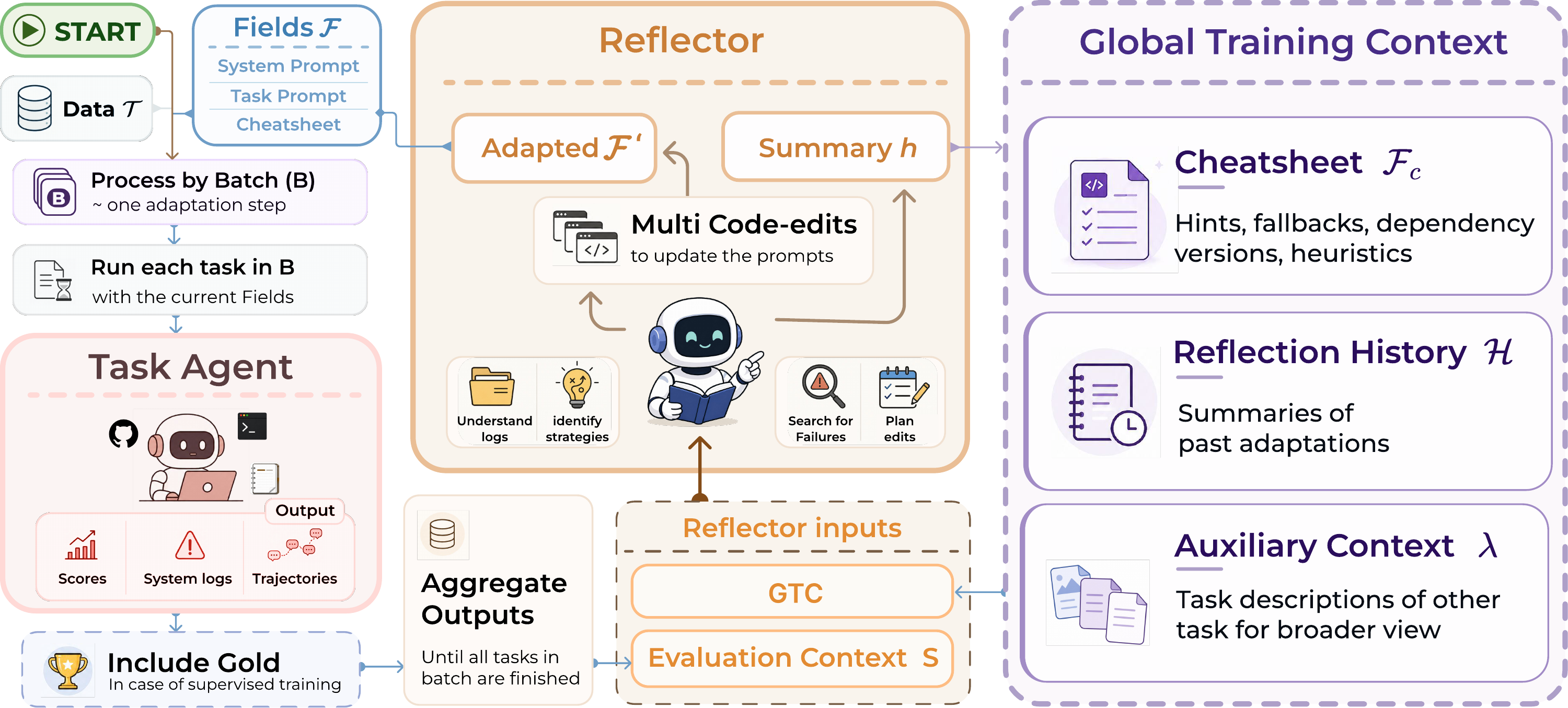}

    \caption{ \textbf{The \sysname\ Framework} 
   A given task agent's prompts, represented as editable Fields (System, Task, Cheatsheet), are optimized by a Reflector agent using a code-edit tool and guided by a Global Training Context and local, per-batch Evaluation Context. 
    }
    \label{fig:arch}
\end{figure*}

To address these gaps, we introduce \sysname\ , a training-free framework for building self-adapting agents tailored to research-coding workflows. \sysname\ adopts a simple, unified design built around three core components: (1) \emph{prompt adaptation over configurable fields}, which defines fields to be optimized and adapts them based on observed failure modes and evaluation feedback; (2) a \emph{Global Training Context} (GTC), which preserves and aggregates experience across tasks and adaptations; and (3) \emph{targeted code-based updates} via a Reflector module, which applies structured edits to prompts and other optimizable fields. Together, these components allow \sysname\ to progressively refine its behavior, reuse prior strategies, and update its reasoning without overfitting to specific tasks. Our work makes the following contributions:
\begin{itemize}
    \item We formulate research code reproduction as a test-time adaptation problem for LLM agents, highlighting concrete failure modes specific to research repositories.

    \item We demonstrate that \sysname\ improves over expert-crafted instructions across three research-coding benchmarks in a fully unsupervised online setting, and further improves when given ground-truth signals, achieving gains of up to $4.50\%$ on SUPER \citep{super}, $1.3\%$ on ResearchCodeBench \citep{rcb}, and $4.89\%$ on ScienceAgentBench \citep{sab}.

    \item We demonstrate that \sysname\ achieves  performance gains over prior prompt adaptation baselines like GEPA \cite{gepa} and ACE \cite{ace} with substantially lower overhead. Adaptation cost is nearly  $10\times$ lower than ACE and GEPA, wall-clock time is up to $2.7\times$ faster.
\end{itemize}

\section{Related Work}

\textbf{Research-Coding benchmarks and approaches:} LLMs are increasingly evaluated across ML engineering benchmarks \citep{mle,mlab}, end-to-end research workflows \citep{heurekabenchbench}, and the broader research experimentation lifecycle \citep{idea2plan,rexbench,paperbench,expbench,abgen}. Benchmarks targeting research-code reproducibility \citep{super,scicode,scireplicatebench,corebench} reveal persistent performance gaps even in multi-agent and search-based systems \citep{paperbench,paper2code,Agentlab,aide,si2026executiongroundedautomatedairesearch}, pointing to the need for agents capable of self-reflection rather than static, handcrafted workflows. While EXP-Bench \citep{expbench} and PaperBench \citep{paperbench} target experiment generation and end-to-end research replication respectively, we focus on SUPER \citep{super}, ResearchCodeBench \citep{rcb}, and ScienceAgentBench \citep{sab}, which capture comparable research-coding challenges with more tractable computational cost and reliable, execution-based evaluation.

\textbf{Prompt optimization and self evolution techniques:} 
Classical prompt optimization treats prompts as tunable parameters via reinforcement learning, gradient-free, or heuristic search~\citep{khattab2023dspycompilingdeclarativelanguage}. Newer approaches introduce reflective and evolutionary strategies~\citep{gepa,mipro}, while runtime-adaptive agents modify their scaffolds on the fly~\citep{zhang2025darwingodelmachineopenended,adas}. Task-level methods~\citep{evomac,aflow} improve instance performance but fail to transfer across tasks. Continual approaches such as CLIN~\citep{majumder2024clin} rely on episodic memory regeneration, while DC~\citep{dynamiccheatsheet} and ACE~\citep{ace} apply structured edits to an evolving memory; however, these methods remain tied to local signals and short-horizon, densely supervised settings. Research coding breaks both assumptions. Tasks are long-horizon, supervision is sparse, and recurring failure patterns emerge only across many executions. \sysname\ addresses this by aggregating 
experience globally and applying code-based edits directly to agent prompts, enabling cumulative adaptation without semantic drift.

\section{Reflective Evolving Research Engineer (\sysname)}
\label{sec:sar}

\subsection{Setup}
\label{subsec:setup}

We formalize the adaptation problem over three editable context fields that govern agent behavior: $\mathcal{F} = \{\mathcal{F}_s, \mathcal{F}_x, \mathcal{F}_c\}$, where $\mathcal{F}_s$ is the \emph{system prompt} (global behavior and rules), $\mathcal{F}_x$ is the \emph{task prompt} (task-specific instructions instantiated at runtime), and $\mathcal{F}_c$ is the \emph{cheatsheet} (a persistent memory, initialized empty, that accumulates reusable strategies and tips). More generally, F can comprise an arbitrary number of user-defined editable fields to suit different agent configurations. Together, these fields parameterize agent behavior without modifying model weights. Given a dataset of tasks $\mathcal{T} = \{(x, o)\}$, where $x$ is a task description and $o$ contains target metrics and optional gold outputs, the agent $\Phi$ produces an output $\hat{o} = \Phi(x; \mathcal{F})$ that is evaluated by a metric function $\mu$ to yield a scalar score $s = \mu(\hat{o}, o)$. Adaptation seeks optimal fields:
\begin{equation}
\mathcal{F}^* = \arg\max_{\mathcal{F}}\; \mathbb{E}_{(x,o)\sim\mathcal{T}}\bigl[\mu\bigl(\Phi(x;\mathcal{F}),\, o\bigr)\bigr].
\label{eq:sysname_objective}
\end{equation}

\subsection{Method Overview}
\label{subsec:overview}
\sysname\ improves a coding agent through an iterative adaptation loop  (Figure~\ref{fig:arch}), progressively editing the three fields of $\mathcal{F}$ based on execution feedback. The agent runs on tasks in batches which provides the Reflector with diverse execution signals, avoiding the diminished feedback that arises when reflecting on all tasks at once without intermediate updates, and amortizing the cost of reflection.  A key component of this loop is the information provided to the Reflector for decision-making: after each batch, it receives a local evaluation signal $S = \{(x, s, \hat{o})\}$, referred to as the \emph{Evaluation Context}, which summarizes batch outcomes and is augmented with ground truth when available, along with a GTC constructed from upcoming task descriptions and prior reflection summaries (Section~\ref{subsec:context}). Together these complement each other to guide the Reflector in diagnosing errors  and making surgical Python-based edits to the three fields. This process repeats across multiple batches, enabling the system to accumulate knowledge over time without rewriting prompts from scratch.
The key mechanism enabling precise adaptation is a \emph{code-based field update}, illustrated in Figure~\ref{fig:codeedits}. Instead of regenerating the full prompt, the Reflector generates a short Python program that modifies only the relevant part of a field. Edits can range from simple string replacements to more complex restructuring, and are run in an isolated environment for safety, and are described in detail in Section~\ref{subsec:cr}. The overall adaptation loop is formalized in Algorithm~\ref{alg:sar}, and the Reflector module in Algorithm~\ref{alg:reflector}.

\noindent
\begin{figure*}[t]
\begin{minipage}[t]{0.52\textwidth}

\begin{algorithm}[H]
\caption{\sysname: Adaptation Process}
\label{alg:sar}
\begin{algorithmic}[1]

\Require Agentic system $\Phi$, Training data $\mathcal{T} = \{(  $Task description $x$, Ground truth $o)\}$,  \\ Metric function $\mu$ , Context fields $\mathcal{F}$

\Ensure Adapted fields $\mathcal{F}$

\For{each epoch}
    \State Shuffle $\mathcal{T}$ and form batches $\{B\}$\
    \Statex \commentq{Initialize Reflection history}
    \State $\mathcal{H} \gets \emptyset$  
    \For{each batch $B$}
    \Statex \commentq{Initialize Evaluation  Context}
        \State $S \gets \emptyset$  
        \ForAll{$(x,o) \in B$}
            \State $\hat{o} \gets \Phi(x,\mathcal{F})$
            \State $s \gets \mu(\hat{o},o)$
            \If{ use gold}
                \State $S \gets S \cup \{(x,s,o,\hat{o})\}$
            \Else
                \State $S \gets S \cup \{(x,s,\hat{o})\}$
            \EndIf
        \EndFor

        \State $\lambda \subseteq \{x \mid (x,o) \in \mathcal{T} - B\}$ 
        \Statex\commentq{Construct Auxiliary context}
        
        \State $(\mathcal{F}',h) \gets$ {\sc Reflector}$(\mathcal{F},S,\mathcal{H},\lambda)$
        \State $\mathcal{H} \gets \mathcal{H} \cup \{h\}$ ;
        $\mathcal{F} \gets \mathcal{F}'$
    \EndFor
\EndFor

\State {\bfseries Output:} Adapted Fields $\mathcal{F}$

\end{algorithmic}
\end{algorithm}

\end{minipage}
\hfill
\begin{minipage}[t]{0.46\textwidth}
\begin{algorithm}[H]
\caption{{\sc Reflector} Agent Module 
}

\label{alg:reflector}
\begin{algorithmic}[1]
\Require 
\Statex Fields $\mathcal{F}=\{\mathcal{F}_s,\mathcal{F}_x,\mathcal{F}_c\}$, LLM $\mathcal{L}$, 
\Statex History buffer $\mathcal{H}$
\Statex Evaluation context $S$ 
\ASpace action a $\in$ \{{\sc Edit},{\sc Finish}\}
\Statex $f' \gets$ {\sc Edit}$(p,f)$:
\Statex \grey{- Program $p$ updates $f \in \{\mathcal{F}_s, \mathcal{F}_x, \mathcal{F}_c\}$ }
\Statex {\sc Finish}$(h)$: 
\Statex \grey{- Terminates the loop with summary $h$}
\State $\mathcal{F}' \gets \mathcal{F}$
\Statex \commentq{Construct reflector memory}
\State $A \gets [S, \mathcal{H},\lambda]$ \Statex \commentq{Reflector loop}

\Repeat
    \Statex \commentq{Action $a$ and it's respective input $c$}
    \State $(a,c) \gets \mathcal{L}(\mathcal{F}',A)$ 
    \If{$a =$ {\sc Edit}}
        \State $(f,p) \gets c$  
        \State $f' \gets$ {\sc Execute}$(p,f)$ 
        \Statex \commentq{Updates appropriate fields}
        \State $\mathcal{F}' \gets f'$ 
        \State  $A \gets A + (a,c)$
    \EndIf
\Until{$a =$ {\sc Finish}}
\Statex \commentq{Generates a summary of changes in $c$}
\State $h \gets c$
\State {\bfseries Output:} $(\mathcal{F}',h)$

\end{algorithmic}

\end{algorithm}
\end{minipage}

\end{figure*}

\subsection{Global Training Context (GTC)}\label{subsec:context}

\sysname\ maintains a Global Training Context that aggregates signals across training iterations, enabling adaptation beyond local feedback via three complementary signals:
 

\begin{enumerate}
    \item \textbf{Cumulative Cheatsheet ($\mathcal{F}_c$)}:  
It is a continually updated, lightweight collection of concise, domain-specific strategies recorded in natural language by the Reflector. Initialized empty, it grows by distilling execution 
experience into actionable reminders used directly by the agent during task execution.
    \item \textbf{Reflection History ($\mathcal{H}$)}: A record of prior reflection summaries, where each entry $h \in \mathcal{H}$ captures the rationale and outcome of an adaptation step. Unlike $\mathcal{F}_c$, which guides 
task execution, $\mathcal{H}$ guides the Reflector itself, by preventing 
contradictory edits from short-term or noisy feedback and promoting 
stable, incremental adaptation across batches.
    \item \textbf{Auxiliary Context ($\lambda$)}: exposes the Reflector to task descriptions beyond the current batch, drawn preferentially from unseen 
training tasks, when no such tasks remain, $\lambda$ is sampled from previously trained task descriptions. This encourages updates that generalize across future task variations rather than overfitting to the current batch.
\end{enumerate}

Together: $\mathcal{F}_c$ provides reusable guidance, $\mathcal{H}$ maintains long-term coherence, and $\lambda$ discourages local optima. Equipped with these signals, the Reflector can make informed field updates.  GTC growth is bounded by design and configurable to fit resource budgets: Auxiliary Context and Reflection History are tunable hyperparameters, and Reflection History resets each epoch (Algorithm 1). Only the Cheatsheet accumulates across iterations. In practice, GTC size remained well within typical context limits throughout our experiments, staying roughly within 20–40k tokens across all three benchmarks.

\subsection{Reflection and Update Mechanism}
\label{subsec:cr}

\begin{figure*}[t]
    \centering
    \resizebox{\textwidth}{!}{
    \includegraphics[width=\textwidth]{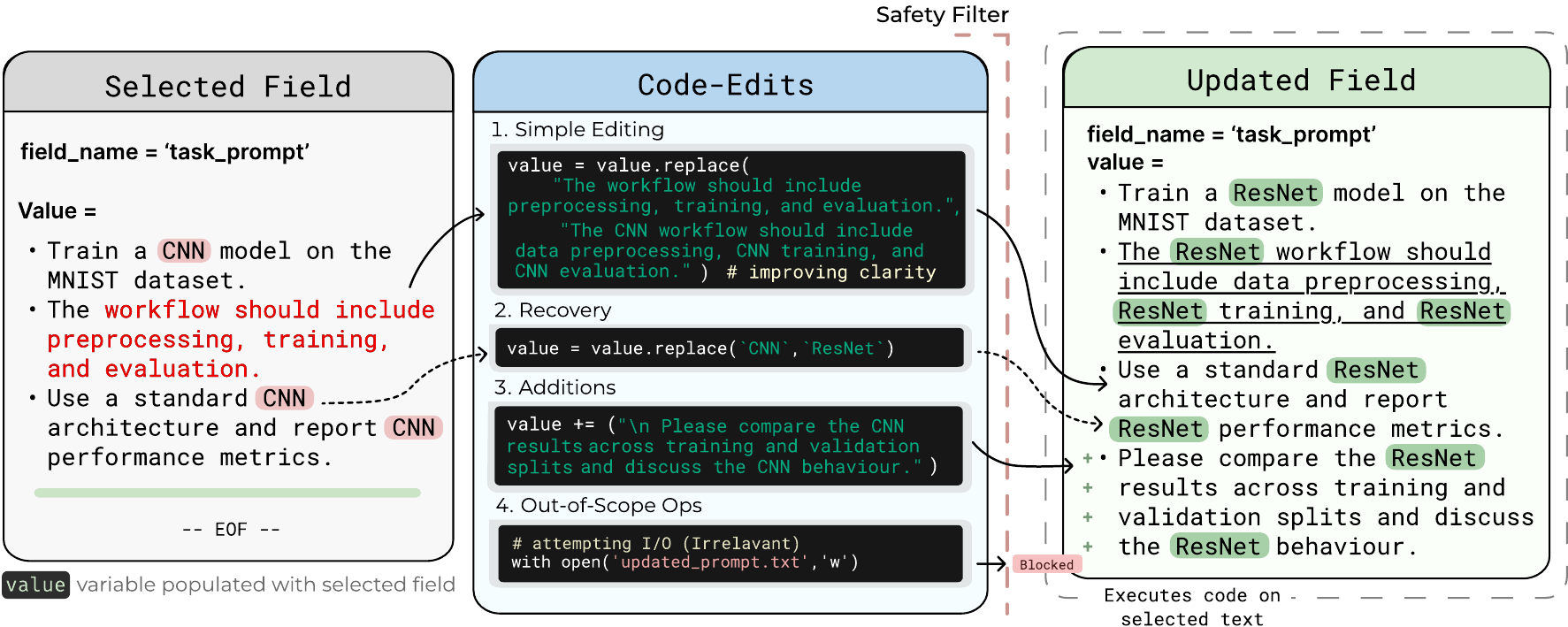}}
    \caption{\textbf{The Reflector code-edit tool}. Targeted Python edits update only the relevant part of a selected field, avoiding the semantic drift of full-prompt rewrites.}
    \label{fig:codeedits}
\end{figure*}

The Reflector is a single agent responsible for both diagnosing failures and editing the fields (Prompt in Appendix~\ref{app:metaprompts}
. Keeping these roles unified --- rather than splitting them across a multi-agent pipeline~\citep{ace, adas} --- preserves a coherent view of the evolving system state and avoids hand-off boundaries that can cause misinterpreted intent and incoherent updates. The central challenge is performing targeted edits without \emph{semantic drift}. Full prompt regeneration tends to silently alter unrelated instructions and overwrite stable, validated content. To address this, we introduce a lightweight \textbf{code-based update tool} inspired by the CodeAct framework~\citep{codeact}. As illustrated in Figure~\ref{fig:codeedits}, the Reflector selects a field $f \in \mathcal{F}$ and generates a short Python program $p$ that modifies it. Given an original task prompt (Figure~\ref{fig:codeedits} left), the Reflector generates Python code (Figure~\ref{fig:codeedits} center) that directly operates on the prompt text, for instance replacing an imprecise instruction with a clearer one, swapping a model reference from CNN to ResNet, or appending new behavioral instructions. Each operation targets only the relevant substring, leaving the rest of the prompt intact. The safety filter intercepts any out-of-scope operations such as file I/O before execution (these restrictions are configurable, allowing users to relax or tighten the safety constraints if needed), and filters them. The approved program is executed in a secure, isolated environment (Figure~\ref{fig:codeedits} right) to produce the updated field ($\mathcal{F}'$). This interface provides three key advantages: 

(i) \textbf{Targeted, low-overhead updates:} Edits are applied only to the relevant portions of $\mathcal{F}$  via code-based transformations, preventing semantic drift and avoiding overwrites of stable content. (ii) \textbf{Expressive, unconstrained modifications:} Unlike template- or rule-based methods \citep{mipro,ace}, Python enables flexible text transformations, leveraging the Reflector's coding ability for precise updates without complex schemas or restrictive APIs. Fields are further authored as Jinja2 templates rather than plain strings, so the Reflector can condition rendered content on task attributes (e.g., domain or task type), reuse shared instruction blocks across conditions, and avoid duplicating near-identical heuristics verbatim — capabilities a flat string field cannot express. (iii) \textbf{Safe, predictable execution:} a two-layer design (static filter + isolated runtime) keeps updates auditable; programs exceeding string operations are rejected and the failure is fed back for retry. While this may cause occasional tool failures, we treat it as a necessary trade-off for  execution safety. The filter is configurable, allowing practitioners to relax constraints for more expressive edits if needed.

\section{Experiment Setup}
\label{sec:expsetup}
We evaluate \sysname\ with a primary focus on long-horizon research-coding tasks.  Additionally, we considered both single-shot and interactive settings to assess robustness and generalization across diverse research workflows (Section~\ref{sec:bench}). For each benchmark, we define offline and online adaptation regimes (Section~\ref{sec:adaptation}) and compare against strong baseline methods (Section~\ref{sec:baselinemeth}). A summary of benchmark datasets, including task counts, approximate per-task inference cost, and resource constraints, is provided in Appendix~\ref{app:exp}.

\subsection{Benchmarks}
\label{sec:bench}
\textbf{SUPER}~\citep{super} is a benchmark of 45 long-horizon research-coding tasks that require agents to set up and execute experiments from real research repositories in a containerized environment. The tasks capture realistic workflows, including dependency installation, configuration, and resolving runtime or version issues. Agent performance is evaluated using the benchmark’s standard metrics: (i) \emph{Output Match} requires reproduced results (e.g., accuracy, F1 score, or error rate) to match expert-reported outputs, (ii) \emph{Landmarks} measure the presence of expected indicators of correct progress in execution logs, with higher scores assigned when more expected signals are observed, and (iii) \emph{Overall} is the average of Output Match and Landmarks, serving as the primary summary metric.

\textbf{ResearchCodeBench (RCB)}~\citep{rcb} evaluates an LLM's ability to re-implement core methodologies from research papers in a \emph{single-shot} setting. For each task, the agent is provided with the paper and partially masked code files, and must reconstruct the missing implementation in a single forward pass. The benchmark comprises 212 tasks from 20 top-tier venues (e.g., ICLR, NeurIPS). Performance is measured by \emph{Accuracy}: each task is scored as pass (1) or fail (0) based on whether the reconstructed code passes hidden unit tests without errors, and Accuracy is the mean score across all tasks.

\textbf{ScienceAgentBench (SAB)}~\citep{sab} evaluates language agents on data-driven scientific tasks in an \emph{interactive} code-generation setting. Each task requires producing a self-contained Python program implementing a component of a scientific workflow, often grounded in machine learning methodologies. Unlike single-shot settings, agents iteratively execute and refine code using runtime feedback. The benchmark contains 102 tasks derived from 44 peer-reviewed publications across four scientific disciplines. Evaluation uses two metrics: (i) Success Rate (SR), measuring whether a program satisfies task-specific execution and output criteria, and (ii) CodeBERTScore (CBS), which measures semantic similarity between generated and reference code using contextual embeddings.

\begin{table*}[!t]
\centering
\small
\resizebox{\textwidth}{!}{
\begin{tabular}{l c | c c c | c | c c }
\toprule
\multirow{2}{*}{\textbf{Method}} &
\multirow{2}{*}{\textbf{GT}} &
\multicolumn{3}{c|}{\textbf{SUPER (\%)}} &
\textbf{RCB (\%)} & \multicolumn{2}{c}{\textbf{SAB (\%)}}\\
 & & \textbf{Landmarks} & \textbf{Output Match} & \textbf{Overall} &
\textbf{Accuracy} & \textbf{SuccessRate} & \textbf{CodeBERTScore} \\ 
\midrule
\rowcolor{gray!8}
 & & \multicolumn{3}{c|}{\emph{long-horizon}} & \emph{single shot} & \multicolumn{2}{c}{\emph{interactive - ReAct} }\\
\midrule
Baseline & --
        & 24.70 
        & 9.30 
        & 17.00
        & 27.80  &20.90 & 67.50  \\

SOTA$^\dagger$& -- 
     & 35.80
     & 14.80
     & 25.30
     & 31.90 & 23.50 & 88.20 \\

\midrule
\rowcolor{gray!8}
\multicolumn{8}{c}{\hspace{2em}\emph{Offline Adaptation}} \\
\midrule

GEPA & \xmark
     & \cvneg{0.00}{35.80}{0.0}
     & \cvpos{18.60}{3.80}{0.85}
     & \cvneg{9.30}{16.00}{0.42}
     & \cvneg{25.69}{6.21}{0.00}
     & \cvpos{17.42}{6.08}{1.35} 
     & \cvneg{77.80}{10.94}{2.33} \\
 
GEPA & \icheckmark
     & \cvneg{1.93}{33.87}{2.15}
     & \cvneg{2.60}{12.20}{1.0}
     & \cvneg{2.27}{23.03}{0.99}
     & \cvneg{10.44}{21.46}{3.12}
     & 19.35\minus{4.15}
     & 65.60\minus{22.60} \\
 
ACE & \xmark
    & \cvneg{19.53}{16.27}{10.01}
    & \cvpos{18.51}{3.71}{3.62}
    & \cvneg{19.02}{6.28}{3.2}
    & \cvneg{21.99}{9.91}{3.28}
    & \cvneg{8.60}{14.90}{7.27}
    & \cvneg{53.48}{34.72}{8.08} \\
 
ACE & \icheckmark
    & \cvneg{27.22}{8.58}{3.04}
    & \cvpos{18.64}{3.84}{3.91}
    & \cvneg{22.93}{2.37}{1.61}
    & \cvneg{7.87}{24.03}{1.75}
    & \cvneg{4.30}{19.20}{1.86}
    & \cvneg{24.28}{63.92}{1.95} \\
 
\sysname\ & \xmark
     & \cvneg{35.67}{0.13}{3.09}
     & \cvpos{19.76}{4.96}{3.71}
     & \cvpos{27.71}{2.41}{2.99}
     & \cvneg{28.75}{3.15}{2.0}
     & \cvneg{23.23}{0.27}{3.14}
     & \cvneg{80.25}{7.95}{2.7} \\
 
\sysname\ & \icheckmark
     &  \cvpos{35.84}{0.04}{5.52}
     & \cvpos{23.76}{8.96}{4.59}
     & \cvpos{29.80}{4.50}{4.26} 
     & \cvpos{33.20}{1.30}{1.80} 
     & \cvpos{28.39}{4.89}{4.36}
     & \cvneg{82.84}{5.36}{1.19} \\
 
\bottomrule
\end{tabular}}
\caption{Offline adaptation Results. Reported values are mean over 5 runs. \icheckmark/\xmark\ denote with/without ground-truth (GT) hints; (–) indicates inference-only.  {\color{modernGreen}$+x$} and {\color{modernRed}$-x$} showing gains and drops relative to Static SOTA results marked by$^\dagger$.}
\label{tab:offline_adapt_results}
\end{table*}

\subsection{Benchmarks Extension for Self-Adaptation}
\label{sec:adaptation}

For each benchmark, we consider two adaptation settings, \emph{offline} and \emph{online}, which differ in task exposure and supervision. In the \textbf{offline} setting, adaptation uses fixed training and validation tasks, while evaluation is performed on a held-out test set, with train/val/test splits of 9/9/27, 34/34/144, and 20/20/62 for SUPER, ResearchCodeBench, and ScienceAgentBench respectively. We evaluate two supervision conditions, denoted \textbf{GT} and \textbf{No-GT}: under GT, the agent receives benchmark-specific ground-truth hints during training (see Appendix~\ref{app:exp} for GT contents); under No-GT, ground truth is withheld entirely and the agent adapts solely from its own execution traces (i.e., unsupervised adaptation). 
In the \textbf{online} setting, tasks arrive sequentially from the full dataset without repetition, and no ground-truth is available at any point, hence inherently \textbf{No-GT}. Since the full task pool includes the tasks used for offline test evaluation, we report both full-dataset and test-subset performance, with the latter enabling direct comparison against the offline test split.

\subsection{Baseline Methods}
\label{sec:baselinemeth}
\textbf{Baseline and SOTA Prompts:}  Our baseline system uses minimal, non-optimized prompts containing only 
the core task description, referred to as `baseline' which serves as a reference for isolating the effects of \sysname's adaptation mechanisms. All adaptation methods, including
REVERE, GEPA, and ACE, start from this baseline prompt. For the SUPER (long-horizon) and SAB (interactive) , we implement a ReAct-style agent \citep{react} equipped with code tools (see Appendix~\ref{app:agent_impl} for details) and for ResearchCodeBench, we used direct LLM calls to generate programs. Additionally, we report the current state-of-the-art performance for each benchmark using author-provided instructions, denoted as `Static SOTA' (see Appendix~\ref{app:staterprompts} for prompts). 

\textbf{GEPA (Genetic-Pareto)} \citep{gepa} is a sample-efficient prompt optimizer that evolves prompts via natural-language reflection and Pareto-based genetic search over prompt candidates with a fixed evaluation pool; its Pareto filter requires scoring all candidates against a held-out validation set. However, this design primarily supports offline optimization, making online adaptation infeasible. 

\textbf{ACE (Agentic Context Engineering)}~\citep{ace} is a playbook-based learning approach inspired by~\citep{dynamiccheatsheet} that maintains bullet-point strategies across predefined categories updated via a multi-agent reflector-curator, processing one task at a time.  We use step size one as recommended in the original paper.

REVERE and ACE adapt using only the training split; GEPA additionally uses the validation split, solely for Pareto candidate selection. See Appendix~\ref{app:hyper} for code and configuration details for \sysname, GEPA, and ACE. All primary results are computed using GPT-4.1, using a single model for both reflection and adaptation to reduce cross-model knowledge transfer. Full model details, including primary and additional models used in analysis with Reflector/task-model configurations, are provided in Appendix~\ref{app:hardware}.

\section{Results and Analysis}
\label{sec:results}

\begin{table*}[!tb]
\centering
\small
\resizebox{\textwidth}{!}{
\begin{tabular}{l |  c c c | c | c c  }
\toprule
\multirow{2}{*}{\textbf{Method}} &
\multicolumn{3}{c|}{\textbf{SUPER Bench (\%)}} &
\textbf{RCB (\%)}  & \multicolumn{2}{c}{\textbf{SAB (\%)}}\\
 & \textbf{Landmarks} & \textbf{Output Match} & \textbf{Overall} &
\textbf{Accuracy} & \textbf{SuccessRate} & \textbf{CodeBERTScore}  \\ 
\midrule

\rowcolor{gray!9}
 &  \multicolumn{3}{c|}{\emph{long-horizon}} & \emph{single shot} & \multicolumn{2}{c}{\emph{interactive} } \\
\midrule

Baseline - Test Tasks & 24.70 & 9.30 & 17.00 & 27.80 & 20.96 & 67.50\\
Baseline - All Tasks & 19.63 & 10.40 & 15.01 & 43.80 & 20.58 & 66.86  \\

\midrule
\rowcolor{gray!7}
\multicolumn{7}{c}{\textit{Online Adaptation}} \\
\midrule

ACE - Test Tasks & 
26.80\plus{2.10} & 11.10\plus{1.80} & 18.95\plus{1.95} & 15.50\minusg{12.30} & 11.30\minusg{9.66} & 66.50\minusg{1.00}  \\
ACE - All Tasks &
24.80\plus{5.17} & 14.20\plus{3.80} & 19.50\plus{4.49} & 19.81\minusg{23.99} & 11.80\minusg{8.78} & 65.5\minusg{1.36} \\

\sysname{} - Test Tasks
& 33.50\plus{8.80} 
& 17.60\plus{8.30} 
& 25.50 \plus{8.10} & 35.40 \plus{7.60} & 25.80 \plus{4.84} & 69.61 \plus{2.11}
\\

\sysname{} - All Tasks
& 30.10\plus{10.40} 
& 18.70\plus{8.30} 
& 24.40\plus{9.39} & 45.00 \plus{1.20} & 25.49 \plus{4.91} & 71.68 \plus{4.82}
\\

\bottomrule
\end{tabular}
} 
\caption{Online Adaptation results. All Tasks:  performance  on full benchmark; Test Tasks: Results on Test set tasks, \plus{x} and \minus{x} with respect to  baseline. 
}
\label{tab:online_adapt_results}
\end{table*}

\begin{table*}[!b]
\centering
\small
\resizebox{\textwidth}{!}{
\begin{tabular}{l | c c c | c | c c }
\toprule
\multirow{2}{*}{\textbf{Method}} & \multicolumn{3}{c}{\textbf{SUPER Bench (\%)}} & \textbf{RCB(\%)}    & \multicolumn{2}{c}{\textbf{SAB (\%)}}\\
& \textbf{Landmarks} &  \textbf{Output Match} & \textbf{Overall} & \textbf{Acc.} & \textbf{SuccessRate} & \textbf{CodeBERTScore} \\ 
\midrule
\rowcolor{gray!10}
& \multicolumn{3}{c|}{\textit{long-horizon}} & \textit{single shot} & \multicolumn{2}{c}{\emph{interactive} } \\
\midrule

\sysname\ 

& 38.10
& 31.40

& 34.70
& 35.40 & 30.60 & 83.3
\\
\sysname\ w / o  Cheatsheet
& 30.70 \minusg{7.40}
& 16.70 \minusg{14.70}
& 23.70 \minusg{11.00}
& 30.60 \minusg{4.80}
& 20.90 \minusg{9.70}
& 78.30 \minusg{5.00}
\\

\sysname\ w / o  Auxiliary context
& 24.80 \minusg{13.30}
& 24.10 \minusg{7.30}
& 24.40 \minusg{10.30}
& 31.90 \minusg{3.50}
& 25.80 \minusg{4.80}
& 79.80 \minusg{3.50}
\\

\sysname\ w / o  Reflection history
& 25.90 \minusg{12.20}
& 9.30 \minusg{22.10}
& 17.60 \minusg{17.10}
& 32.60 \minusg{2.80}
& 22.50 \minusg{8.10}
& 75.90 \minusg{7.40}
\\

\sysname\ w / o  GTC
& 21.70 \minusg{16.40}
& 10.20 \minusg{21.20}
& 16.00 \minusg{18.70}
& 27.10 \minusg{8.30}
& 19.30 \minusg{11.30}
& 77.30 \minusg{6.00}
\\

\sysname\ w / o  Code Edits
& 0.00 \minusg{38.10}
& 1.11 \minusg{30.29} 

& 0.55 \minusg{34.15} 
& 26.38 \minusg{9.02} 
& 4.83 \minusg{25.77} 
& 59.88 \minusg{23.42}
\\

\bottomrule
\end{tabular}
} 
\caption{Ablation on Offline Adaptation. Results correspond to a representative run, \minusg{x} indicates difference with respect to \sysname\ (full model)}
\label{tab:offline_ablation_results}
\end{table*}

\textbf{Under offline adaptation, \sysname\ is the only method that consistently benefits from stronger supervision signals, improving over Static SOTA across all benchmarks and metrics.} Results are presented in Table \ref{tab:offline_adapt_results}. On SUPER, \sysname\ improves both Output Match and Landmarks simultaneously, while GEPA improves Output Match at the cost of Landmarks collapsing to zero -- a direct consequence of its discard-and-rewrite design overfitting recent traces and discarding stable reasoning structure. ACE shows marginal, inconsistent gains that remain below \sysname\ throughout.

\textbf{\sysname's global aggregation of signals lets it keep improving on domain-heavy benchmarks where GEPA and ACE degrade under stronger supervision.} The more revealing pattern emerges on ResearchCodeBench and ScienceAgentBench, which are domain-knowledge-heavy with limited structural overlap across examples, and thus constrained by the amount of transferable signal any adaptation method can exploit. Here, GEPA and ACE both degrade further when given gold hints: GEPA's rewrite mechanism discards valuable trajectory supervision, while ACE indiscriminately absorbs step-level signals that do not transfer across heterogeneous tasks, which is evident in how their prompts evolve under adaptation (Appendix \ref{app:prompt-growth} Figure~\ref{fig:adapted_prompts}). \sysname\ instead aggregates signals globally, allowing the Reflector to distinguish recurring cross-task patterns from instance-level noise, then distributing learned strategies across the system prompt, task prompt, and cheatsheet rather than collapsing all adaptation pressure onto a single field. This makes \sysname\ the only method whose performance remains stable and consistently improves even when supervision is inherently noisy and task-specific 
(see Appendix~\ref{app:offline_stats} for standard deviations and significance tests). REVERE shows similar robustness to initialization quality beyond the
baseline prompt (see Appendix~\ref{app:cold-start}).

\textbf{\sysname's gains come from learning to solve tasks correctly, not from matching a reference answer.} On ScienceAgentBench, \sysname\ achieves CodeBERTScore closest to SOTA among all methods while improving SuccessRate by +4.89\% over SOTA. This gap reflects a known limitation of CodeBERTScore: lexical similarity to a single reference implementation does not necessarily track functional correctness, making SuccessRate the more meaningful indicator of task success in this interactive setting (see Appendix~\ref{app:sab-cbs-sr} for a task-level reconciliation, including matched examples where \sysname\ achieves SR=1 with lower CBS due to functionally valid alternative implementations).
We provide an additional case study analyzing autonomously discovered failure modes of SUPER baseline agent and prompt evolution across adaptation iterations in Appendix~\ref{app:auto_disc} 
and Appendix~\ref{app:super_prompt_evo}.

\textbf{\sysname\ retains comparable performance in online settings despite minimal feedback.} Table~\ref{tab:online_adapt_results} reports the stricter online regime, where tasks arrive sequentially, each encountered once, and no ground-truth labels are available at any point. \sysname\ improves over the baseline across all benchmarks and metrics, while ACE frequently falls below the baseline, indicating that a cheatsheet updated from local signals alone is insufficient for stable cross-task generalization without the broader context that \sysname's GTC provides. Lower absolute performance relative to offline training is structurally expected: processed tasks cannot be revisited, future tasks are unseen at adaptation time, and supervision is entirely absent. Yet relative gains remain consistent across benchmarks, confirming that \sysname's adaptation is driven by recurring execution patterns rather than ground-truth dependence. Notably, on ResearchCodeBench, \sysname\ surpasses its own offline result, partly due to the online setting including a broader task pool with partially overlapping test tasks that provide additional adaptation signal. On SUPER, \sysname\ online matches Static SOTA without any ground-truth access, underscoring that effective prompt evolution does not require supervision to approach expert-crafted performance. We further examine REVERE's sensitivity to model and Reflector choice in
Appendix~D.1.

\begin{figure*}[!t]
    \centering
    \includegraphics[width=\textwidth]{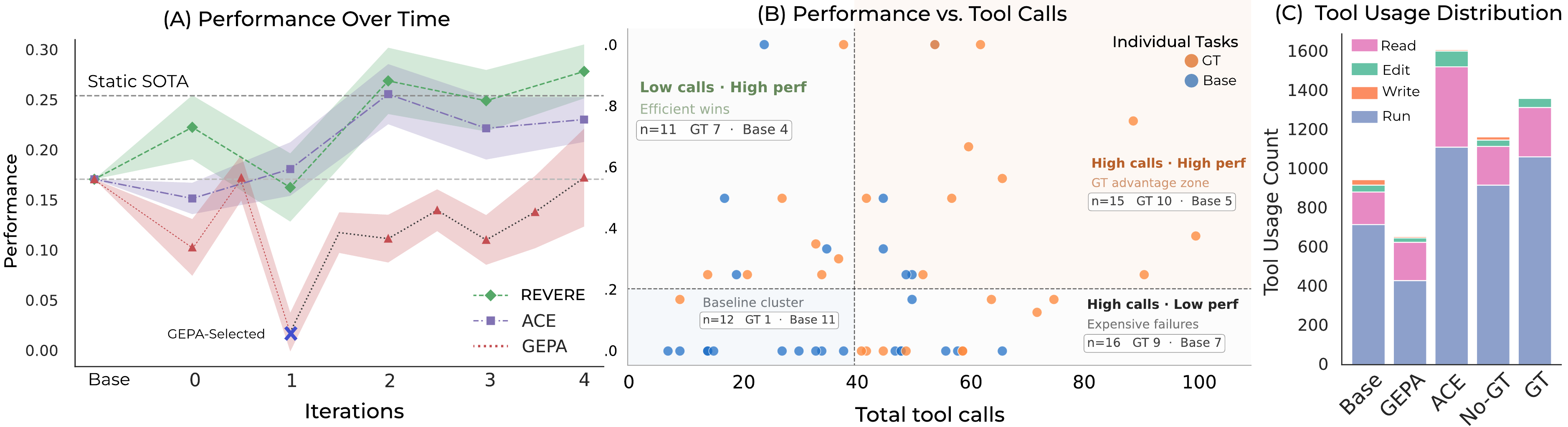}
    
\caption{\textbf{Performance Analysis on the adapted agent on SUPER Overall metric.} 
\textbf{(A)} Performance trajectory across iterations. 
\textbf{(B)} Per-task performance vs. total tool calls; orange = \sysname\ (GT), blue = Baseline.
\textbf{(C)} Tool usage across variants (Baseline, GEPA, ACE, \sysname\ w/o GT, \sysname\ w/ GT). 
\textit{Note:} GEPA iterations denote total candidates and are normalized to the axis scale; the blue marker indicates the Pareto-selected prompt.}
    \label{fig:performance}
\end{figure*}

\textbf{Do we need all components of \sysname?} The ablation results (Table~\ref{tab:offline_ablation_results}) confirm that \sysname’s effectiveness arises from the interaction of its core components. Removing the cheatsheet memory causes broad degradation across metrics, showing that cumulative retention of discovered fixes and heuristics is central to adaptation. Excluding auxiliary context or reflection history individually degrades performance, but for different reasons: auxiliary context lets the Reflector anticipate patterns in upcoming tasks, while reflection history prevents past mistakes from recurring. The largest drops occur when the entire GTC is removed, confirming these forward- and backward-looking signals are complementary. Most strikingly, replacing targeted code-based edits with full-prompt regeneration collapses performance almost entirely, a larger drop than removing the GTC itself (Table~\ref{tab:offline_ablation_results}). This confirms the two components play distinct roles: the GTC drives what gets learned, while the code-edit mechanism is what sustains those updates across iterations. Without it, accumulated signal is overwritten before it can meaningfully learn (a retention analysis following this failure mode is included in Appendix~\ref{app:retention}).

\textbf{\sysname\ improves steadily across iterations instead of locking early on weak adaptation.}
Figure~\ref{fig:performance}(A) shows \sysname\ improving steadily across iterations, recovering from a marginal early dip to surpass both Static SOTA and ACE by the final iteration. GEPA, in contrast, is unable to capture the heterogeneity of long-horizon tasks or account for learning across iterations, causing its heuristic-based Pareto selection to lock in a weaker prompt based on early evaluation, highlighting the fragility of heuristic-based selection without accumulated adaptation context. \textbf{An intuitive learning curve emerges under adaptation, spanning efficient wins, persistent successes, and informative failures.}
Figure~\ref{fig:performance}(B) reveals that baseline tasks cluster in the bottom-left reflecting shallow exploration, while \sysname-GT tasks spread across all quadrants, indicating that adaptation induces richer exploratory behaviour. This dispersion captures three distinct regimes: efficient wins (top-left) where tasks are resolved with minimal tool usage, persistent successes (top-right) where harder tasks require greater investment, and informative failures (bottom-right) where extended traces provide the richest adaptation signal for future improvement. Together these dimensions show that exploration depth, efficiency, and persistence are all actively shaped by \sysname's learning process.   A detailed trajectory-level breakdown of improving, declining, and neutral tasks provided in Appendix~\ref{app:corr_toolvsperf}.

\textbf{More tool calls alone don't drive progress — what matters is converting them into stronger gains.}
Figure~\ref{fig:performance}(C) reveals how adaptation strategy shapes tool usage. GEPA under-explores relative to the baseline, consistent with its context-discarding design. Both ACE and \sysname\ show increased tool usage over the baseline, consistent with their performance gains. However \sysname\ converts tool calls into stronger performance gains more efficiently than ACE, confirming that our method adapts the underlying agent efficiently.

\begin{figure*}[!t]
    \centering
    \includegraphics[width=\textwidth]{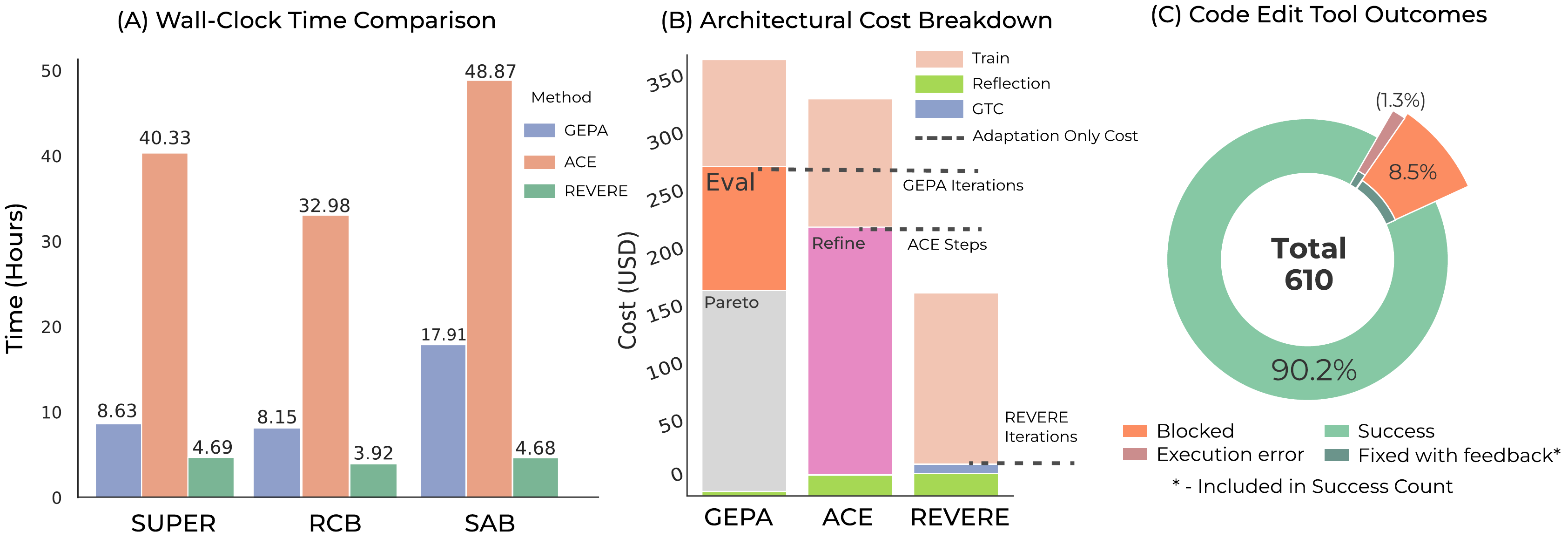}

    \caption{\textbf{Efficiency analysis of \sysname\ } \textbf{(A)} Illustrates wall-clock time for offline adaptation of \sysname\ with baselines. \textbf{(B)} Compares computational cost (USD) on the SUPER task in an offline setting across \sysname, ACE, and GEPA, decomposed by architectural components. Dashed lines indicate sole adaptation cost. \textbf{(C)} Distribution of outcomes across code-edit tool calls in the ground-truth setting across all benchmarks.}
    \label{fig:efficiency}
\end{figure*}

\textbf{Speed and cost scale in \sysname's favor: adaptation stays lightweight and cost-effective without sacrificing reliability.} Figure~\ref{fig:efficiency}(A) shows \sysname\ completes adaptation in a fraction of the wall-clock time of ACE and GEPA across all benchmarks, over 2.7x faster, aided by parallelised execution and efficient memory management. Figure~\ref{fig:efficiency}(B) compares adaptation costs (component breakdown in Appendix~\ref{app:cost}). Training cost (shown hatched) is shared across all methods and thus not a meaningful differentiator; curation costs are also negligible (\sysname: \$0.16 vs. ACE: \$0.47 on SUPER). The key difference is adaptation overhead: \sysname's adaptation-only cost is nearly 10x lower than ACE and GEPA, which incur repeated per-task refinement steps and Pareto re-evaluation respectively, both scaling unfavorably with task complexity. 
\sysname\ avoids this through targeted edits and a compact GTC, keeping per-task overhead low even as context grows (see Appendix~\ref{app:des_rati} for cost accounting rationale).

\textbf{Most code-based edits succeed outright, and the safety filter turns the rest into a practical, self-correcting safeguard rather than a bottleneck.} Figure~\ref{fig:efficiency}(C) reports that 90.2\% of code-based edit attempts succeed outright; remaining cases are caught by the safety filter, which is configurable. We recommend retaining it, as the Reflector self-corrects blocked edits within the same iteration, making this a practical safeguard for safe LLM-to-LLM integration.

\section{Conclusion}

We present \sysname, a lightweight, self-adapting framework for LLM agents, applied especially to research-coding tasks. \sysname\ maintains a Global Training Context -- reflection history, auxiliary context, and a cumulative cheatsheet -- alongside local, per-batch evaluation signals, and applies precise code-edits to prompts, more efficiently than existing prompt updating mechanisms. This supports both offline and online adaptation while remaining interpretable and easy to integrate. REVERE shows consistent gains over strong baselines (GEPA, ACE) across three settings — SUPER (long-horizon), ResearchCodeBench (single-shot), and ScienceAgentBench (interactive) — under both supervised and unsupervised learning, while being nearly 10× more cost-effective and 2.7× faster than prior methods. Our analysis shows gains are more modest on domain-heavy benchmarks, suggesting prompt adaptation alone may not capture task-specific knowledge, and that the growing GTC may accumulate stale information over time. Exploring task-specific adaptation and GTC pruning are natural directions for future work.

%% file: appendix.tex
\appendix

\onecolumn


\section{ Experiment Details }





\subsection{Benchmark datasets and resource limits }
\label{app:exp}

Table~\ref{tab:dataset} reports task counts, task inputs, GT content, per-task inference cost, and resource limits (LLM calls and combined agent + evaluation time) across the three benchmarks. Costs reflect single-run agent execution and vary with task complexity and trajectory length. SUPER is the most expensive due to its long-horizon interactive nature; we use the Expert subset, which contains expert-designed tasks and is therefore more challenging and representative of real-world research scenarios. ResearchCodeBench is the most lightweight given its single-shot setting. Resource limits are fixed and held constant across all methods to ensure a fair and controlled comparison. GT content is deliberately kept incomplete across all three benchmarks, allowing hints to guide adaptation without letting agents shortcut the task by copying the answer outright. For SUPER, the expert trajectory comes from a structurally different execution environment and omits exact dependency versions, which are unreliable to reproduce and known to affect baselines inconsistently depending on when they are run. For ResearchCodeBench, only partially masked code is given, since the full implementation would leak answers across tasks sharing masking patterns. For ScienceAgentBench, gold execution results and outputs are withheld so agents cannot infer expected outputs without actually implementing and reasoning through the task. This isolates genuine adaptation gains from direct answer-copying, keeping the comparison across methods rigorous.

\begin{table*}[h!]
\centering
\resizebox{\textwidth}{!}{
\begin{tabular}{l c l l c c} \\
\midrule
\textbf{Benchmark} & \textbf{Tasks} & \textbf{Task Input} & \textbf{Ground Truth} & \textbf{Cost P.T} & \makecell{\textbf{Resource Limits} \\ \textbf{(Calls / Exec Time)}} \\ 
\midrule
\textbf{Super-Expert} & 45 & Query, GitHub repo, commit & Expert trajectory & \$1--3 & 100 iters / 20 min \\
\multicolumn{6}{l}{
\textit{Setting up, and executing tasks from research repositories.}
} \\
\textbf{ResearchCodeBench} & 212 & Paper, masked code, context files & Target code & $\sim$ \$0.2 & 1 Call / 10 min \\
\multicolumn{6}{l}{
\textit{Code completion by implementing methodology from the paper.}
} \\
\textbf{ScienceAgentBench} & 102 & Task instruction, dataset path & Gold Python file & $\sim$ \$0.3 & 20 iters / 40 min \\
\multicolumn{6}{l}{
\textit{Scientific coding tasks spanning multiple domains for data-driven discovery.}
} \\
\bottomrule
\end{tabular}}
\caption{Benchmark datasets used in our evaluation, with task counts, task inputs, ground-truth (GT) content provided to REVERE under the GT-assisted offline setting, approximate per-task inference cost (Cost P.T), and per-task resource limits set for each task (no. of LLM calls and time for agent + evaluation execution). Costs are estimates based on gpt-4.1.}
\label{tab:dataset}
\end{table*}

\subsection{Implementation Details}
\label{app:hyper}

\paragraph{REVERE.}
For \sysname, training was performed for 5 epochs with a batch size of 3 on 
SUPER, a batch size of 7 on ScienceAgentBench, and a batch size 5 on ResearchCodeBench. Auxiliary context sizes were set to 6 (SUPER), 3 (ResearchCodeBench, reduced due to longer code and paper context segments), and 5 (ScienceAgentBench). No separate validation set was used for REVERE, allowing all available training data to contribute directly to 
parameter updates. In the online setting, the auxiliary context $\lambda$ is sampled from previously 
encountered tasks only, as no future tasks are accessible in the online setting. 


\paragraph{GEPA.}
We use GEPA's official implementation~\footnote{\url{https://github.com/gepa-ai/gepa-artifact}} 
with 32 optimization iterations on SUPER (early stopping with 20 consecutive 
iterations without improvement, mini feedback set of 3 examples, Pareto 
frontier size 9), a budget of 600 metric evaluations ($\approx$20 iterations) 
on ResearchCodeBench, and a maximum of 15 iterations with mini feedback set of size 7 on 
ScienceAgentBench. GEPA additionally uses the held-out validation split solely for Pareto candidate selection. For fair cost comparison in Figure~\ref{fig:efficiency}, GEPA's cost and wall-clock time analysis are capped to appropriate iterations to match the same training sample count as \sysname \ although it ran for more iterations with no improvements.

\paragraph{ACE.}
We use ACE's official implementation~\footnote{\url{https://github.com/ace-agent/ace}} 
with a step size of 1 (i.e., batch size 1) across all benchmarks, consistent 
with the original ACE paper's implementation~\cite{ace}.

Refer to Appendix~\ref{app:cost} for 
architectural breakdown of ACE and GEPA methods.

\subsection{Agentic and Evaluation Implementation}
\label{app:agent_impl}

Since REVERE, ACE, and GEPA operate under architecturally distinct frameworks, directly comparing them requires a shared foundation. To this end, we standardize the baseline agent, tool interfaces, and execution environment across all three methods, ensuring that observed performance differences reflect genuine methodological distinctions rather than implementation artifacts. We implement a ReAct-style agent that interleaves reasoning and tool usage to solve complex coding and research tasks through an iterative interaction loop with the environment.  The agent is built on a modular tool-calling framework using our custom-built \texttt{aicodetools}\footnote{\url{https://pypi.org/project/aicodetools/}}, inspired by prior agent frameworks such as InspectAI~\citep{inspectaiuk} and LangChain. Our library abstracts system-level complexities and isolates execution overhead from the adaptation framework, allowing efficient extensibility and integration of new capabilities. It supports containerized execution and parallel task handling, ensuring scalability, reproducibility, and efficient resource utilization.
The agent is equipped with core tools including \textbf{read}, \textbf{write}, \textbf{edit}, and \textbf{run\_command}, enabling controlled interaction with the filesystem and execution environment. Additionally, we introduce a \textbf{finish} tool that allows the agent to produce structured outputs, including final answers, execution summaries, and explicit reasoning traces, thereby improving interpretability and evaluation consistency.

We further extend the evaluation benchmarks---SUPER, Research Code Bench (RCB), and ScienceAgentBench---by integrating Dockerized execution, a streamlined pipeline, and enhanced evaluation mechanisms. Critically, because all methods share the same containerized environment, any misconfiguration is immediately visible and affects all methods uniformly, eliminating environment-specific confounds as a source of variance. These contributions ensure safe and reproducible execution while enabling efficient experimentation through parallelization and standardized logging. In addition, \sysname\ framework is designed to be highly extensible, supporting seamless incorporation of additional benchmarks, tools, and evaluation protocols, and thereby facilitating continued development and broader applicability of agent-based research systems.

\subsection{System Hardware}
\label{app:hardware}
All methods including \sysname\ and baselines use the Azure \texttt{gpt-4.1} (api version : \texttt{2025-03-01-preview}) model and all experiments are conducted on \texttt{standard\_d16ads\_v5} with 16 vCPUs and 64 GiB memory. For further analysis in Appendix~\ref{app:model-heterogeneity}, we evaluate heterogeneous Reflector/task-model configurations spanning GPT-5.4, GPT-5.4-mini, GPT-5.2, and \texttt{gemini-3-flash-preview} from Google AI Studio.

We select GPT-4.1 as our primary model for its 1M-token context window, needed to handle the extreme context requirements of offline adaptation — inputs of 300k--500k tokens for the Reflector and 40k--120k tokens for agents, driven by large code repositories, academic papers, and long-horizon reasoning traces. The online setting carries lower context requirements per adaptation step, which is why Appendix~\ref{app:model-heterogeneity} additionally evaluates smaller-context models (GPT-5.4-mini, GPT-5.2) alongside GPT-4.1, Gemini-3-Flash.

\section{Robustness on Adaptation }
\label{app:offline_stats}
\begin{table*}[!t]
\centering
\small

\resizebox{\textwidth}{!}{
\begin{tabular}{l c | c c c | c | c c }
\toprule
\multirow{2}{*}{\textbf{Method}} &
\multirow{2}{*}{\textbf{GT}} &
\multicolumn{3}{c|}{\textbf{SUPER Bench (\%)}} &
\textbf{RCB (\%)} & \multicolumn{2}{c}{\textbf{ScienceAgentBench (\%)}}\\
 & & \textbf{Landmarks} & \textbf{Output Match} & \textbf{Overall} &
\textbf{Accuracy} & \textbf{SuccessRate} & \textbf{CodeBERTScore} \\ 
\midrule
\rowcolor{gray!8}
 & & \multicolumn{3}{c|}{\emph{long-horizon}} & \emph{single shot} & \multicolumn{2}{c}{\emph{interactive - ReAct} }\\
\midrule
\rowcolor{gray!8}
\multicolumn{8}{c}{\hspace{2em}\emph{Offline Adaptation}} \\
\midrule

GEPA & \xmark
     & \cvnegv{0.0}{35.80}{0.0}
     & \cvposv{18.6}{3.80}{0.85}
     & \cvnegv{9.30}{16.00}{0.42}
     & \cvnegv{25.69}{6.21}{0.00}
     & \cvposv{17.42}{6.08}{1.35} 
     & \cvnegv{77.80}{10.94}{2.33}  \\

GEPA & \icheckmark
     & \cvnegv{1.93}{33.87}{2.15}
     & \cvnegv{2.60}{12.20}{1.0}
     & \cvnegv{2.27}{23.03}{0.99}
     & \cvnegv{10.44}{21.46}{3.12}
      &\cvnegv{17.09}{6.41}{2.70} & \cvnegv{78.53}{9.67}{1.84}\\

ACE & \xmark
    & \cvnegv{19.53}{16.27}{10.01}
    & \cvposv{18.51}{3.71}{3.62}
    & \cvnegv{19.02}{6.28}{3.2}
    & \cvnegv{21.99}{9.91}{3.28}
    & \cvnegv{8.60}{14.90}{7.27}
    & \cvnegv{53.48}{34.72}{8.08} \\

ACE & \icheckmark
    & \cvnegv{27.22}{8.58}{3.04}
    & \cvposv{18.64}{3.84}{3.91}
    & \cvnegv{22.93}{2.37}{1.61}
    & \cvnegv{7.87}{24.03}{1.75}
    & \cvnegv{4.30}{19.20}{1.86}
    & \cvnegv{24.28}{63.92}{1.95} \\

\sysname\ & \xmark
     & \cvnegv{35.67}{0.13}{3.09}
     & \cvposv{19.76}{4.96}{3.71}
     & \cvposv{27.71}{2.41}{2.99}
     & \cvnegv{28.75}{3.15}{2.0}
     & \cvnegv{23.23}{0.27}{3.14}
     & \cvnegv{80.25}{7.95}{2.7} \\

\sysname\ & \icheckmark
     & \cvposv{35.84}{0.04}{5.52}
     & \cvposv{23.76}{8.96}{4.59}
     & \cvposv{29.80}{4.50}{4.26}
     & \cvposv{33.20}{1.30}{1.8} 
     & \cvposv{28.39}{4.89}{4.36}
     & \cvnegv{82.84}{5.36}{1.19} \\

\bottomrule
\end{tabular}
}
\caption{Offline Adaptation Results; RCB: ResearchCodeBench; with (\icheckmark) and without (\xmark) ground-truth (GT) Hints, $\pm$: standard deviation (Std Dev) across 5 runs.}
\label{tab:offline_stats}
\end{table*}

\begin{figure*}[t]
    \centering
    \includegraphics[width=1\textwidth]{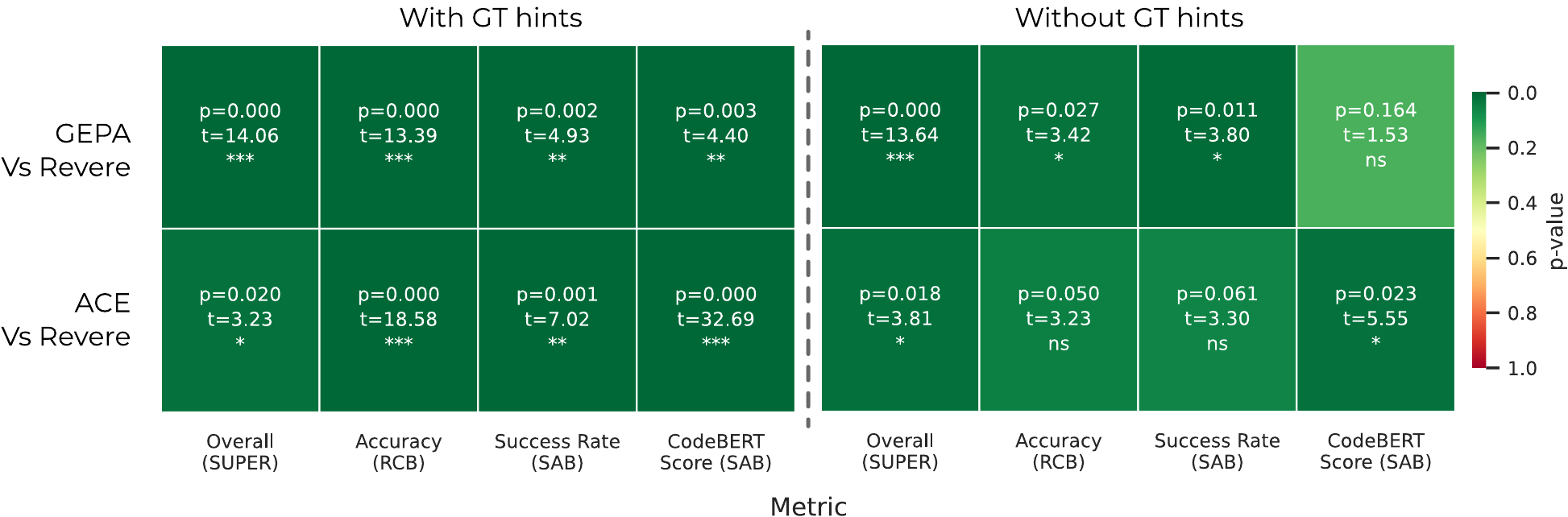}

    \caption{ Welch t-test : \sysname\ ***$ p < 0.001$, ** $p < 0.01$ ; * $p < 0.05$ ; $ns$ \: $p \ge 0.05$ ; $t > 0$ \text{ means \sysname\ has higher mean}\textbf{}
 }
    \label{fig:ttest}
\end{figure*}

\begin{figure*}[t]
    \centering
    \includegraphics[width=1\textwidth]{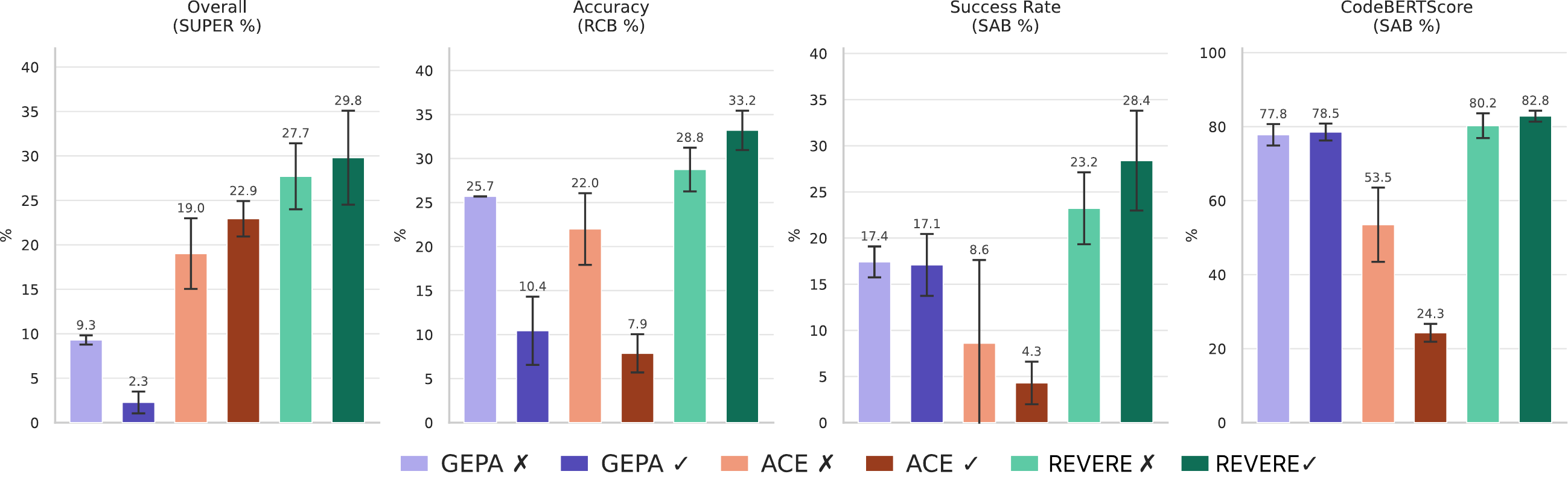}

    \caption{ Offline Adaptation score with 95\% $CI$, where GT \icheckmark = Adapted with ground-truth hints; \xmark = Adapted with-out ground-truth hints;
 }
    \label{fig:ci}
\end{figure*}

Table~\ref{tab:offline_stats} reports mean performance and standard deviations across five runs. Figure~\ref{fig:ttest} and Figure~\ref{fig:ci} provide statistical support for the results discussed in Section~\ref{sec:results}.

\textbf{Significance of gains with ground-truth supervision.} With GT hints, \sysname\ achieves statistically significant improvements over both GEPA and ACE across all benchmarks and metrics ($p < 0.001$, $t > 0$). The high $t$-statistics in this setting, particularly on RCB Accuracy ($t = 13.39$ vs. GEPA, $t = 18.58$ vs. ACE) and SAB CodeBERTScore ($t = 4.40$ vs. GEPA, $t = 32.69$ vs. ACE), reflect that \sysname\ not only achieves higher means but does so consistently across runs, as supported by the standard deviations in Table~\ref{tab:offline_stats}.

\textbf{Significance without ground-truth supervision.} Without GT hints, significance is largely retained on SUPER . However, significance drops on domain-knowledge-heavy benchmarks, particularly SAB, where \sysname\ vs. ACE yields $p = 0.050$ on Accuracy and $p = 0.061$ on Success Rate. This is expected: adapting to tasks that require deep domain-specific knowledge from execution traces alone is a hard problem. Importantly, \sysname\ still maintains higher means throughout ($t > 0$), indicating a consistent directional advantage even where significance is not achieved.

\textbf{Consistency of gains across runs.} Figure~\ref{fig:ci} shows 95\% confidence intervals across all methods and metrics. \sysname\ improves consistently over competing methods, with little to no overlap between its confidence intervals and those of GEPA and ACE on primary metrics. This is most visible on SUPER Overall and SAB Success Rate, where \sysname's lower confidence bound exceeds the upper bound of competing methods in several comparisons. The overlap that does occur is limited to domain-knowledge-heavy settings without GT supervision, which is consistent with the significance analysis above.

\section{Prompt Adaptation Analysis}
\label{app:prompt-growth}

\begin{figure*}[t]
    \centering
    \includegraphics[width=0.9\textwidth]{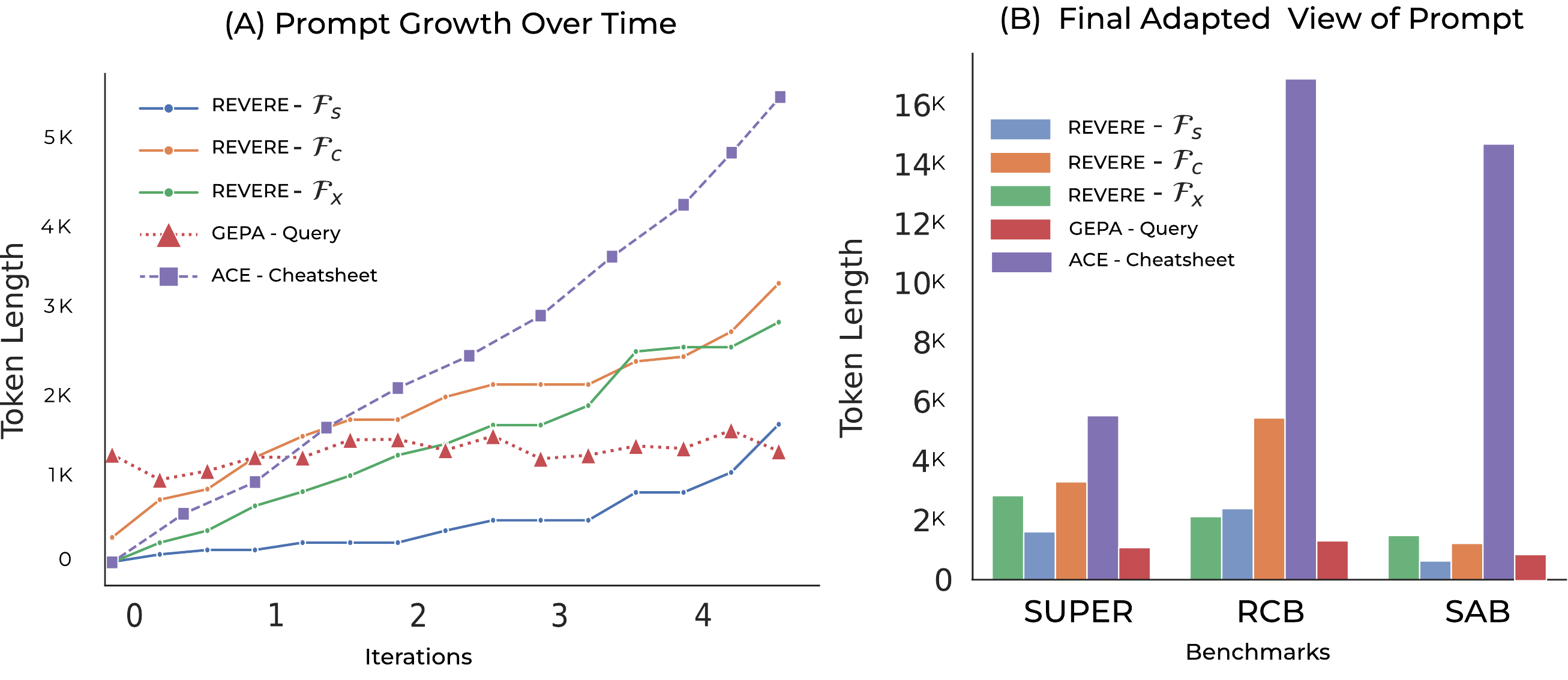}

    \caption{ \textbf{Prompt Analysis}: \textbf{(A)} Illustrates prompt component growth in token length across iterations on the SUPER benchmark, comparing \sysname\ components , GEPA query, and ACE cheatsheet. \textbf{(B)} Illustrates final adapted prompt lengths (in tokens) across benchmarks, used in offline adaptation results with ground truth. 
 }
    \label{fig:adapted_prompts}
\end{figure*}

\paragraph{\sysname\ maintains efficient adaptation with controlled prompt growth.}
Figure~\ref{fig:adapted_prompts} (A) shows that the three methods diverge sharply in how they manage prompt length across iterations, and the growth patterns reflect their respective architectural designs. \sysname's components ($\mathcal{F}_s$, $\mathcal{F}_c$, $\mathcal{F}_x$) exhibit controlled, distributed growth, allowing the system to allocate capacity where it is most needed. ACE shows steep linear growth driven entirely by its cheatsheet, conflating task-specific and generalizable signals into bloated prompts; growing context length remains a fundamental challenge even in \sysname, and we discuss potential mitigation in Appendix~\ref{app:context_growth}. GEPA maintains consistently low token count by regenerating instructions from scratch each iteration, but discards previously accumulated knowledge at every step, making it poorly suited for domain-rich, individually complex tasks. 
Figure~\ref{fig:adapted_prompts} (B) reports final adapted prompt lengths for the results in Table~\ref{tab:offline_adapt_results}, and directly supports our earlier observation on domain-heavy benchmarks. ACE's cheatsheet grows dramatically on RCB, reflecting the benchmark's heavy reliance on specialized domain knowledge that is difficult to compress into reusable prompts, accumulating task-specific traces instead of distilling generalizable signals. \sysname\ maintains compact prompts across all benchmarks, though as noted earlier, controlled prompt growth alone does not overcome the domain familiarity constraints that bound performance on research-oriented coding tasks.

\subsection{Semantic Drift and Edit Operation Analysis}
\label{app:retention}

\begin{table}[ht]
\centering
\resizebox{\textwidth}{!}{
\begin{tabular}{lccc}
\toprule
Method & Levenshtein Retention & Git Retention & Semantic Similarity \\
\midrule
GEPA -- full prompt rewrite      & 0.11 & 0.03 & 0.62 \\
ACE -- incremental delta updates & 0.86 & 0.92 & 0.99 \\
REVERE -- targeted edits         & 0.84 & 0.89 & 0.94 \\
\bottomrule
\end{tabular}}
\caption{Prompt retention across consecutive adaptation iterations on SUPER. GEPA's
near-zero retention reflects the semantic drift from full-prompt rewriting. ACE's
near-perfect retention reflects its append-only design: new content is structurally and
semantically similar to existing content, so little changes between iterations. REVERE's
intermediate retention ($\sim$0.85--0.95) reflects targeted restructuring: stable content
is preserved while only relevant portions are modified.}
\label{tab:retention}
\end{table}

To directly evaluate whether targeted code-based edits reduce semantic drift relative to
full-prompt rewriting, we measure prompt retention across consecutive adaptation
iterations using three metrics: Levenshtein retention ($1 -$ normalized Levenshtein
distance between prompt versions), git-diff retention ($1 -$ fraction of lines changed
per diff between consecutive versions), and semantic similarity (cosine similarity via
sentence-transformer embeddings, averaged across all edit steps). Higher values indicate
greater similarity to the previous iteration, with 1.0 denoting no change.

ACE's high retention does not necessarily indicate better adaptation. It mainly reflects accumulation, with similar instructions repeatedly added, leading to the unbounded prompt growth shown in
Figure~\ref{fig:adapted_prompts}A. To confirm that REVERE's retention score reflects genuine restructuring rather than
passive accumulation, we analyze the percentage of code-based tool calls using each
Python string operation (a single tool call may invoke multiple operations), reported
for the full model and its ablation variants in Table~\ref{tab:opbreakdown}. Dominant
use of \texttt{split}/\texttt{join} (63.1\%, 62.7\%) and \texttt{insert} (45.9\%),
against low \texttt{replace} (7.5\%) and accumulation ops \texttt{append}/\texttt{+=}
(6.0\%, 14.2\%), indicates structural reorganization rather than wholesale rewriting
(replace-dominant, as in GEPA) or pure accumulation (append-dominant, as in ACE). This
pattern holds across ablation variants, confirming it reflects the edit mechanism
itself rather than any single GTC component.

\begin{table}[ht]
\centering
\resizebox{\linewidth}{!}{
\begin{tabular}{lccccccccccccc}
\toprule
Method & split & join & replace & append & extend & insert & += & strip & startswith & endswith & find & index & count \\
\midrule
REVERE (full)           & 63.1 & 62.7 & 7.5  & 6.0  & 0.0 & 45.9 & 14.2 & 29.9 & 18.7 & 0.7 & 6.7 & 3.4 & 0.0 \\
w/o GTC                 & 75.7 & 73.5 & 4.3  & 14.3 & 2.6 & 53.5 & 11.7 & 24.8 & 12.6 & 0.9 & 3.0 & 9.6 & 0.0 \\
w/o Auxiliary context   & 59.9 & 58.3 & 11.5 & 12.0 & 0.5 & 34.4 & 11.5 & 32.8 & 14.1 & 4.2 & 1.6 & 3.6 & 0.5 \\
w/o Reflection history  & 65.9 & 65.5 & 3.7  & 6.4  & 0.4 & 51.3 & 16.5 & 33.0 & 19.9 & 0.7 & 4.1 & 4.5 & 0.0 \\
\bottomrule
\end{tabular}}
\caption{Percentage of code-based edits using each Python string operation, for REVERE
(full model) and its ablation variants (SUPER, ground-truth setting). The diverse
operation mix -- spanning restructuring (\texttt{split}, \texttt{join}), targeted
correction (\texttt{replace}), and expansion (\texttt{insert}, \texttt{append}) --
confirms genuine restructuring rather than simple accumulation, and this pattern holds
across ablations, indicating it is not an artifact of any single GTC component. For
reference, GEPA's full rewrites are functionally equivalent to 100\% \texttt{replace}
operations, and ACE's append-only curator uses 100\% \texttt{append}/\texttt{+=}
operations. \texttt{w/o Cheatsheet} is omitted, as it only removes the cheatsheet field
at test time and does not alter the training-time edit mechanism.}
\label{tab:opbreakdown}
\end{table}

\section{Analysis on Research Reproduction and Adaptation}
\label{app:case}

\subsection{Generalization Across Model Configurations}
\label{app:model-heterogeneity}

\begin{table}[t]
\centering
\resizebox{\textwidth}{!}{
\begin{tabular}{l|ll |c| c c}
\toprule
Row & Task Model & Reflector & Same/Cross & Test Subset (27) & Full Set (45) \\
\midrule
\midrule
1 & GPT-5.4        & (unadapted) & -- & 48.40 & 47.56 \\
2 & GPT-5.4-mini  & (unadapted) & -- & 41.99 & 43.27 \\
3 & Gemini-3-Flash& (unadapted) & -- & 38.07 & 37.09 \\
\midrule
4 & GPT-5.4-mini & GPT-5.4-mini   & Same  & 33.66 & 39.12 \\
5 & GPT-5.4-mini & GPT-5.4        & Cross & 51.48 & 48.85 \\
6 & GPT-5.4-mini & GPT-5.2        & Cross & 50.00 & 47.07 \\
\midrule
7 & Gemini-3-Flash & Gemini-3-Flash & Same & 57.20 & 53.21 \\
\midrule
8  & GPT-5.4 & GPT-5.4-mini    & Cross & 48.41 & 48.30 \\
9  & GPT-5.4 & GPT-5.2         & Cross & 54.33 & 55.49 \\
10 & GPT-5.4 & Gemini-3-Flash  & \textbf{Cross-family} & 55.20 & 54.71 \\
11 & GPT-5.4 & GPT-5.4         & Same  & 56.90 & 59.29 \\
\midrule
\bottomrule
\end{tabular}}
\caption{Heterogeneous Reflector/task-model configurations under offline
adaptation. The Same/Cross column indicates whether the Reflector and
task model belong to the same model family. Rows 1--3 report unadapted
baselines for each task model. Rows 4--6 isolate Reflector strength on a
weak task model (GPT-5.4-mini). Row 7 reports the same-family
Gemini-3-Flash configuration. Rows 8--11 isolate Reflector strength on a
strong task model (GPT-5.4); row 10 (\textbf{bolded}) pairs GPT-5.4 as
the task model with Gemini-3-Flash as the Reflector, a cross-family
configuration where the task model and Reflector come from different
AI labs.}
\label{tab:reflector-heterogeneity}
\end{table}

Table~\ref{tab:reflector-heterogeneity} reports offline adaptation results under heterogeneous Reflector and task-model configurations, extending the single-model design (Section~4.3) to isolate whether REVERE's gains stem from the adaptation mechanism itself or from the underlying model's inherent capability. Pairing a weaker task model (GPT-5.4-mini) with a stronger Reflector (GPT-5.4 or GPT-5.2; rows 5, 6) surpasses the same task model's unadapted baseline (row 2) and, on the full task set, approaches the unadapted strong-model baseline (row 1) -- despite the task model itself being unchanged. This indicates that adaptation gains are driven predominantly by the quality of reflection rather than task-model capability alone. Conversely, pairing a strong task model with a weak Reflector (row 8) yields only marginal improvement over its unadapted baseline, confirming that Reflector quality is the more critical factor in determining final performance. This pattern also holds within a single family: Gemini-3-Flash under same-family adaptation (row 7) improves substantially over its own unadapted baseline (row 3), the largest relative gain observed in Table~\ref{tab:reflector-heterogeneity}.

To test generalization beyond a single model family, we additionally evaluate a fully cross-family configuration: GPT-5.4 as the task model with Gemini-3-Flash-Preview as the Reflector (row 10). This configuration achieves 55.20\% / 54.71\% (Test Subset / Full Set), comparable to the same-family GPT-5.4 + GPT-5.2 pairing (row 9) and only marginally below the matched GPT-5.4 + GPT-5.4 ceiling (row 11). Row 7 further confirms this ceiling holds under a same-family Gemini-3-Flash pairing. Together, these results show that REVERE's adaptation mechanism transfers across model families and is not tied to same Reflector/task-model pairings, though absolute performance and reflection quality remain sensitive to the underlying models' individual capabilities.

\subsection{Autonomous Failure Discovery from Super Baseline}
\label{app:auto_disc}

\begin{table*}[ht]
\centering
\small
\begin{tabular}{p{0.16\textwidth}p{0.26\textwidth}p{0.52\textwidth}}
\toprule
\textbf{Category} & \textbf{Description} & \textbf{Found Solution} \\
\midrule

Dependencies &
Build/install failures or missing libraries for training/fine-tuning. &
\begin{minipage}[t]{\linewidth}
    \textbullet\ Use \texttt{pip install --only-binary=:all: ...}\\
    \textbullet\ Pin \texttt{numpy}/\texttt{sklearn}/\texttt{Cython} to highest available wheel
\end{minipage}
\\
\midrule

Environment &
Environment/version mismatches causing runtime errors. &
\begin{minipage}[t]{\linewidth}
    \textbullet\ Enumerate/validate environment variables and dependency availability \\
    \textbullet\ Debug versions/signatures: \texttt{inspect.signature(func)} \\
    \textbullet\ Reconfigure environment: if no gpu available configure \texttt{cuda()} to \texttt{cpu}
\end{minipage}
\\
\midrule

Configuration &
Wrong CLI args/configs or missing scripts. &
\begin{minipage}[t]{\linewidth}
    \textbullet\ Verify argument names: \texttt{train\_file}, \texttt{validation\_file}, \texttt{data\_dir}\\
    \textbullet\ Enumerate \texttt{run\_*.sh} candidates and match closest
\end{minipage}
\\
\midrule

Data - Acquisition &
Missing datasets or incorrect asset content. &
\begin{minipage}[t]{\linewidth}
    \textbullet\ Inspect README/scripts; attempt \texttt{wget}/\texttt{gdown}/\texttt{curl}\\
    \textbullet\ Inspect content directly using \texttt{head}, \texttt{cat}, \texttt{less}, \texttt{file}\\
\end{minipage}
\\
\midrule

Data - Preprocessing &
Incompatible dataset schema/fields or container formats. &
\begin{minipage}[t]{\linewidth}
    \textbullet\ Write wrapper to rename or transform fields\\
    \textbullet\ Decompress \texttt{.csv}; check for embedded \texttt{.jsonl}/\texttt{.tsv}
\end{minipage}
\\
\midrule

Execution Issues &
Scripts produce no metrics or import errors. &
\begin{minipage}[t]{\linewidth}
    \textbullet\ Check output directory and config paths\\
    \textbullet\ Fix hardcoded outputs\\
    \textbullet\ Patch imports/\texttt{sys.path}; test package and script entrypoints
\end{minipage}
\\
\midrule

Goal &
Metric extraction and schema compliance for outputs. &
\begin{minipage}[t]{\linewidth}
    \textbullet\ Enumerate \texttt{result.txt} per run; extract requested metrics\\
    \textbullet\ Extract available final metrics from logs/stdout\\
    \textbullet\ Set required keys to null only after full recovery attempts
\end{minipage}
\\
\midrule

Miscellaneous &
Reproducibility and source-of-truth guidance. &
\begin{minipage}[t]{\linewidth}
    \textbullet\ Enumerate \texttt{instance\_id}, query, \texttt{github\_repo}, \texttt{git\_commit}\\
    \textbullet\ Use README/scripts as canonical hyperparameters
\end{minipage}
\\
\midrule

Semantics &
Output semantics: schema matching and null policies. &
\begin{minipage}[t]{\linewidth}
    \textbullet\ Match output schema exactly to the query\\
    \textbullet\ Return null only after exhaustive dataset recovery attempts
\end{minipage}
\\
\bottomrule
\end{tabular}
\caption{Autonomously identified Failures and Solutions of base prompt across research repositories by \sysname\ .}
\label{tab:auto_failures}
\end{table*}

A key observation in Table \ref{tab:auto_failures} is that the dominant failure modes are not algorithmic or model-reasoning errors, but infrastructure and workflow mismatches: dependency resolution, environment configuration, script selection, data formatting, and metric extraction. These issues arise from discrepancies between implicit assumptions held by repository authors and the explicit instructions available to the agent. Crucially, these mismatches recur across different repositories with similar patterns, indicating that research-code reproduction exhibits a shared latent structure of operational pitfalls. This explains why static prompts or local prompt rewrites struggle: each individual failure appears repository-specific, but the type of failure (e.g., missing assets, version drift, schema mismatch) is globally recurring. \sysname’s advantage stems from converting these superficially idiosyncratic errors into reusable procedural heuristics. The cheatsheet thus functions as a cross-repository “operational prior,” allowing the agent to approach new tasks with expectations about likely breakdown points rather than treating each environment as independent.

\subsection{Prompt evolution}
\label{app:super_prompt_evo}

\begin{figure*}[t]
    \centering
    \includegraphics[width=0.9\textwidth]{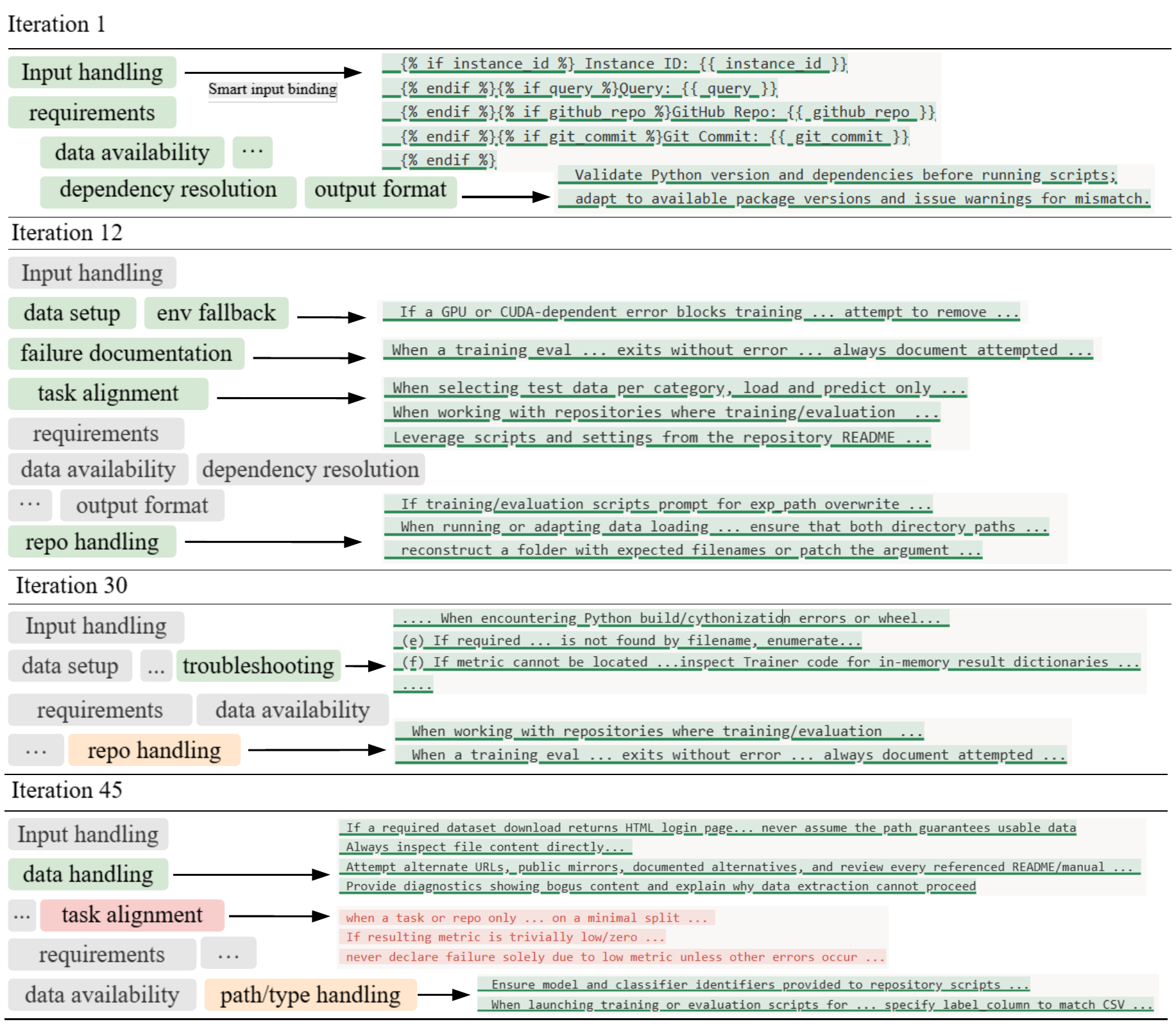}

    \caption{ Evolution of the prompt $\mathcal{F}_x$ across online adaptation iterations. Snapshots from the first, twelfth, thirtieth, and final iterations illustrate how semantically grouped instruction capsules are incrementally refined. Additions (green), modifications (orange), and removals (red) show that adaptation proceeds via structured accumulation rather than prompt rewriting.
 }
    \label{fig:trained_query_flow}
\end{figure*}
Figure \ref{fig:trained_query_flow} visualizes the evolution of the task prompt $\mathcal{F}_x$ across online adaptation iterations, showing incremental additions, modifications, and removals of semantically grouped instruction units. Several patterns emerge. First, growth is structured rather than expansive: instructions are added in coherent clusters tied to concrete failure classes (e.g., environment validation, data inspection, metric recovery), rather than as diffuse prompt length increases. Second, many edits refine existing rules instead of replacing them, indicating that adaptation operates through policy sharpening rather than wholesale behavioral shifts. Third, deletions are rare and localized, suggesting that once a heuristic proves broadly useful, it remains stable across tasks. Together, these dynamics support the claim that \sysname’s improvement mechanism is cumulative and non-destructive: knowledge about workflow regularities is preserved and layered, enabling generalization without the prompt drift or knowledge loss typical of rewrite-based optimization. This explains the empirical stability observed across tasks and the balanced metric gains reported in Section \ref{sec:results}.

\subsection{Cold-Start Robustness}
\label{app:cold-start}

All results in the main paper adapt from a weak, non-expert baseline
prompt (Appendix~\ref{app:staterprompts}), rather than from Static SOTA. To further test
whether REVERE depends on a reasonably-designed starting point, we
evaluate online adaptation from a counter-instructed prompt (see Appendix \ref{app:prompt-cold}) whose
instructions directly contradict the behaviors encoded in Static SOTA, using GPT-5.4 on SUPER.

\begin{table}[t]
\centering
\resizebox{\textwidth}{!}{
\begin{tabular}{llcccccc}
\toprule
Adaptation & Start Prompt & \multicolumn{3}{c}{Full Set (45 Tasks)} & \multicolumn{3}{c}{Test Set (27 Tasks)} \\
 & & Output Match & Landmarks & Overall & Output Match & Landmarks & Overall \\
\midrule
Unadapted & Baseline            & 42.60 & 52.50 & 47.60 & 40.70 & 56.00 & 48.40 \\
Unadapted & Counter-instructed  & 35.10 & 44.70 & 39.90 & 34.40 & 45.20 & 39.80 \\
REVERE & Baseline            & 46.90 & 71.70 & 59.30 & 46.80 & 67.00 & 56.90 \\
REVERE & Counter-instructed  & 43.70 & 68.30 & 56.00 & 41.20 & 69.60 & 55.40 \\
\bottomrule
\end{tabular}}
\caption{Online adaptation from a counter-instructed starting prompt vs.
the standard baseline prompt, on SUPER using GPT-5.4.}
\label{tab:cold-start}
\end{table}

Adapting from the counter-instructed prompt, REVERE reaches 56.00\%
Overall, recovering most of the gap to standard-prompt adaptation
(59.30\%) and substantially exceeding the unadapted standard baseline
(47.60\%). This indicates that REVERE's gains do not depend on a
reasonably-designed initial prompt: even when starting from instructions
that actively contradict expert-crafted guidance, online adaptation
converges toward comparable performance. A small residual gap (3 points)
persists, indicating that initialization quality has a minor but
non-negligible effect on final performance.

\subsection{SUPER with LLM-as-Judge Metric}
\label{app:corr_toolvsperf}
\begin{figure*}[t]
    \centering
    \includegraphics[width=\textwidth]{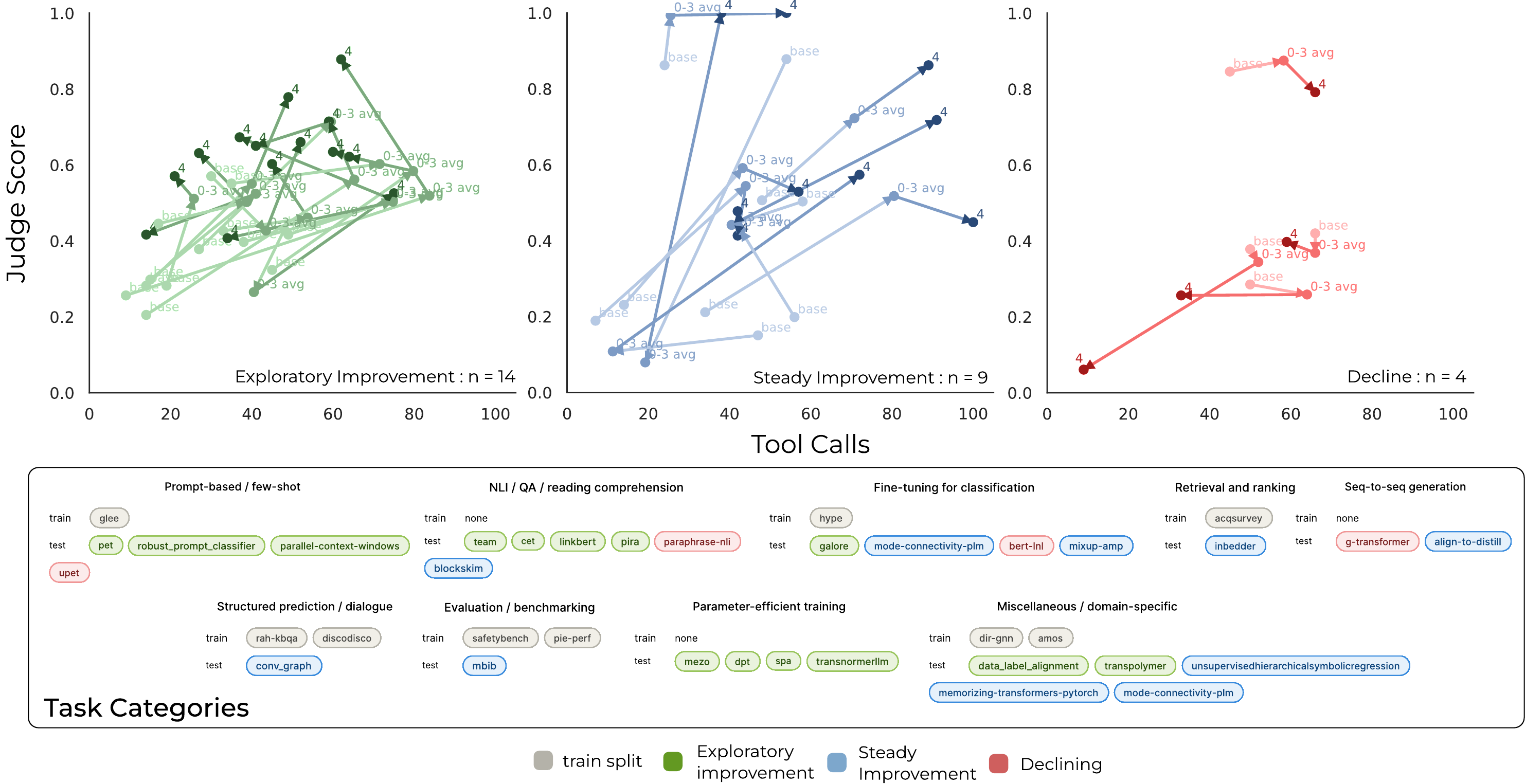}

    \caption{ Offline adaptation trajectory per task on the SUPER benchmark (Judge Score). Each point represents a single test task plotted against its total tool calls. Labels denote adaptation stage: \texttt{base} = baseline (pre-adaptation), \texttt{0--3 avg} = average Judge Score across epochs 0--3, and \texttt{4} = final adapted score at epoch 4. Arrows trace the trajectory of each task from baseline through adaptation. Tasks are colored by their outcome group: \textcolor{modernGreen}{green} = exploratory improvement, \textcolor{modernRed}{red} = declining, \textcolor{blue}{blue} = steady improvement; gray points denote training-split tasks. The bottom panel shows the methodological domain of each task (e.g., parameter-efficient training, NLI/QA, prompt-based), with test tasks highlighted in their outcome color to reveal which domain clusters benefit from adaptation and which do not.
 }
    \label{fig:task_evolve}
\end{figure*}

As the SUPER metric does not track intermediate progress within a task pipeline, we introduce an LLM-as-judge rubric inspired by the autonomous failure modes identified in SUPER (Table~\ref{tab:auto_failures}). This rubric acts as a common evaluation policy across tasks, capturing partial credit for correctly completed pipeline stages even when downstream steps fail. We then define a composite score:

$$\text{JudgeScore} = \frac{\text{LLM Rubric Score} + \text{Overall Metric}}{2}$$
such that the original SUPER metric remains present in the final evaluation, while the judge component surfaces progress that the binary pipeline metric would otherwise obscure.

The test tasks span following broad methodological domains: fine-tuning for classification, prompt-based and few-shot learning, parameter-efficient training, and sequence-to-sequence generation, with a smaller set covering structured prediction, retrieval, and domain-specific tasks.

The largest cluster of 14 exploratory improving tasks is drawn almost entirely from the parameter-efficient training (mezo, dpt, spa, transnormerllm), NLI/QA (team, cet, linkbert, pira), and prompt-based (pet, robust\_prompt\_classifier, parallel-context-windows) domains. Within these, several tasks share direct methodological overlap with training tasks: galore with hype, pet and parallel-context-windows with glee, and team and cet with rah-kbqa. The graph shows a characteristic arc for this cluster, an initial exploratory phase with high tool call counts and moderate scores, followed by convergence toward higher scores at lower tool usage, suggesting the agent generalised how to navigate a class of pipeline from training rather than anything task-specific.

The steady improvement cluster of 9 tasks shows incremental gains over baseline, with several tasks maintaining performance comparable to the base model while using fewer tool calls. Structurally, these tasks cluster within or adjacent to training task domains, particularly around NLI/QA and fine-tuning neighbourhoods shared with rah-kbqa, hype, and glee. This domain proximity appears sufficient to prevent regression but not enough to drive the broader generalisation seen in the exploratory cluster, producing a pattern of bounded, conservative improvement.
The declining cluster of 4 tasks (paraphrase-nli, g-transformer, bert-lnl, upet) shows limited domain overlap with the training set. bert-lnl and upet share their domains with training tasks hype and glee respectively, yet both declined or remained unchanged, highlighting that domain familiarity alone is not sufficient and that other factors govern learnability. bert-lnl already achieved the highest baseline score in the cluster and its adapted score remained the same in the overall metric. paraphrase-nli and g-transformer have no training analogue, and both declined. Both have explicitly multi-stage pipelines by their task descriptions and failure at an earlier stage compounds over the final score regardless of what the agent does correctly in the subsequent stage, which is particularly consequential in a research code execution setting where pipeline steps are tightly coupled.

\section{Discussion}

\subsection{Context Length Growth in Prompt-Based Adaptation}
\label{app:context_growth}
Growing context length is a fundamental challenge in prompt-based adaptation. ACE~\citep{ace} identifies this limitation, noting that aggressive pruning risks discarding relevant context and that advances in long-context modeling are a natural path toward resolving it. We partially mitigate this by limiting the number of additions the Reflector can make per training batch. Two complementary directions offer further relief:
\begin{itemize}
\item \textbf{Periodic flushing with recovery.} For settings where clusters of similar tasks recur, timely flushing of stale content with recovery capability keeps prompts tractable without permanent information loss.
\item \textbf{Adaptive frequency reduction.} For in-domain settings where adaptation stabilizes, reducing adaptation frequency as edit sizes shrink toward zero, analogous to learning rate annealing, offers a principled stopping criterion as the prompt approaches saturation.
\end{itemize}

\subsection{Cost Decomposition of Adaptation Methods}
\label{app:cost}

\textbf{GEPA} is inspired by genetic mutation, iteratively evolving a candidate pool of prompts through selection, mutation, and evaluation. The pipeline proceeds as follows:

\begin{itemize}
\item \textbf{Candidate Pool} and \textbf{Pareto Filter.} GEPA maintains a pool of prompt candidates across iterations. Each candidate in the pool has a corresponding score on every task in the Pareto validation set, which is used to track and compare candidate quality. At each iteration, the best performing candidate is selected from the pool using a Pareto filter over the validation set.
\item \textbf{Minibatch Evaluation (Train).} The selected prompt is evaluated against a minibatch of tasks to assess current performance.
\item \textbf{Reflective Prompt Mutation (Reflection).} Based on execution feedback, a new candidate instruction is proposed via reflection.
\item \textbf{Re-evaluation (Eval).} The mutated prompt is evaluated again on the same minibatch. If performance does not improve, the candidate is discarded, making this evaluation cost entirely wasted.
\item \textbf{Pool Addition or Discard.} If the mutated prompt improves on the minibatch, it proceeds to the next step. Otherwise it is discarded.
\item \textbf{Pareto Addition.} The surviving candidate is scored against all tasks in the full Pareto validation set before being added to the pool. This is computationally expensive regardless of how much the candidate ultimately improves, and represents a significant overhead in long-horizon settings.
\end{itemize}

\textbf{ACE} follows a multi-agent pipeline inspired by generation, reflection, refinement, and curation, processing one task at a time. The pipeline proceeds as follows:
\begin{itemize}
\item \textbf{Generator.} The agent executes the current task using the existing prompt, producing a trajectory.
\item \textbf{Reflector.} After execution, the Reflector analyzes the trajectory and generates metadata keys capturing what succeeded and what failed.
\item \textbf{Refinement.} In cases of failure, the agent re-attempts the task through multiple runs until it succeeds. While effective in short-horizon interactive settings, refinement can be redundant in long-horizon tasks where the agent progresses incrementally and repeated re-attempts risk making the system overly task-specific.
\item \textbf{Curator.} Using the reflected metadata, the Curator edits the prompt, appending new signals to the cheatsheet.
\end{itemize}

In Figure~\ref{fig:efficiency}(C), for GEPA : minibatch evaluation is reported as \textit{Train}, re-evaluation as \textit{Eval}, reflective prompt mutation as \textit{Reflection}, and Pareto scoring as \textit{Pareto}. For ACE : generator execution is reported as \textit{Train}, reflection as \textit{Reflection}, refinement as \textit{Refine}, and curation as \textit{ACE Steps}.

\subsection{Design Rationale for Avoiding Refinement Overhead}
\label{app:des_rati}
\textbf{\sysname\ } is designed to avoid the overhead introduced by refinement and Pareto selection. As LLM reasoning and instruction-following capabilities continue to improve, execution-free evaluation of prompt quality becomes increasingly reliable: a sufficiently capable model with access to the GTC can distinguish effective from ineffective prompts without requiring held-out validation, reducing the need for expensive Pareto re-evaluation. Similarly, access to accumulated failure context across batches makes explicit re-attempts redundant, since the Reflector already has a reasoned account of what failed and why before issuing any edit. The cost of maintaining and querying the GTC appears in the comparison as the reflection component, and represents a principled tradeoff: a moderate, structured overhead in place of the heavier and less predictable costs of refinement and candidate evaluation.

\subsection{ScienceAgentBench: Why CodeBERTScore Drops While Success Rate Improves}
\label{app:sab-cbs-sr}

By the benchmark's scoring rule, any program that passes (SR $=1$) is automatically given a
CodeBERTScore (CBS) of 1.0. This means the CBS average reported in Table~\ref{tab:offline_adapt_results}
is mostly determined by the programs that \emph{fail} (SR $=0$), not by how similar the
passing programs are to the reference code. So a drop in average CBS does not necessarily mean
REVERE's code looks less like the reference --- it can also happen simply because REVERE
passes more tasks that a weaker method would have failed (and scored near-zero on).

To separate these two effects, we recompute CBS using only the actual similarity scores for
programs that pass (excluding the automatic 1.0):

\begin{table}[ht]
\centering
\begin{tabular}{lcc}
\toprule
Metric & Static SOTA & REVERE \\
\midrule
Avg. CBS on passing programs only & 84.2 & 77.0 \\
Programs that fail to run (of 102) & 19 & 9 \\
\bottomrule
\end{tabular}
\caption{CodeBERTScore computed only on programs that pass (SR $=1$), and count of programs that fail to execute, on ScienceAgentBench.}
\label{tab:sab-cbs-sr}
\end{table}

REVERE cuts the number of failing programs roughly in half ($19 \rightarrow 9$), which explains its higher Success Rate. Among programs that pass, REVERE's code still scores lower on CBS than Static SOTA's --- not because it copies or guesses the expected output (the agent never sees reference outputs and has no control over execution), but because it sometimes reaches a working solution structured differently from the single reference file CBS compares against.

\paragraph{Example (Task 7) from benchmark.} Both the reference solution and REVERE's solution pass (SR $=1$), but they are written differently: the reference selects a fixed set of columns, REVERE detects columns automatically; the reference plots a heatmap with 2 subplots, REVERE plots a single bar
chart; the reference assumes an exact data schema, REVERE checks and adapts to the schema at runtime. Both approaches solve the task correctly, but CBS scores them as dissimilar because it measures closeness to one specific reference file, not whether the code works. We see the same pattern in 10 other passing tasks with CBS between 74.3--76.6\% (Task IDs 38, 51, 57, 79--81,
88, 93, 96, 100).

\clearpage

\section{prompts} 
\subsection{Baseline and STATIC-SOTA prompts}
\label{app:staterprompts}
\begin{figure*}[!h]
    \centering
    \includegraphics[width=0.9\textwidth]{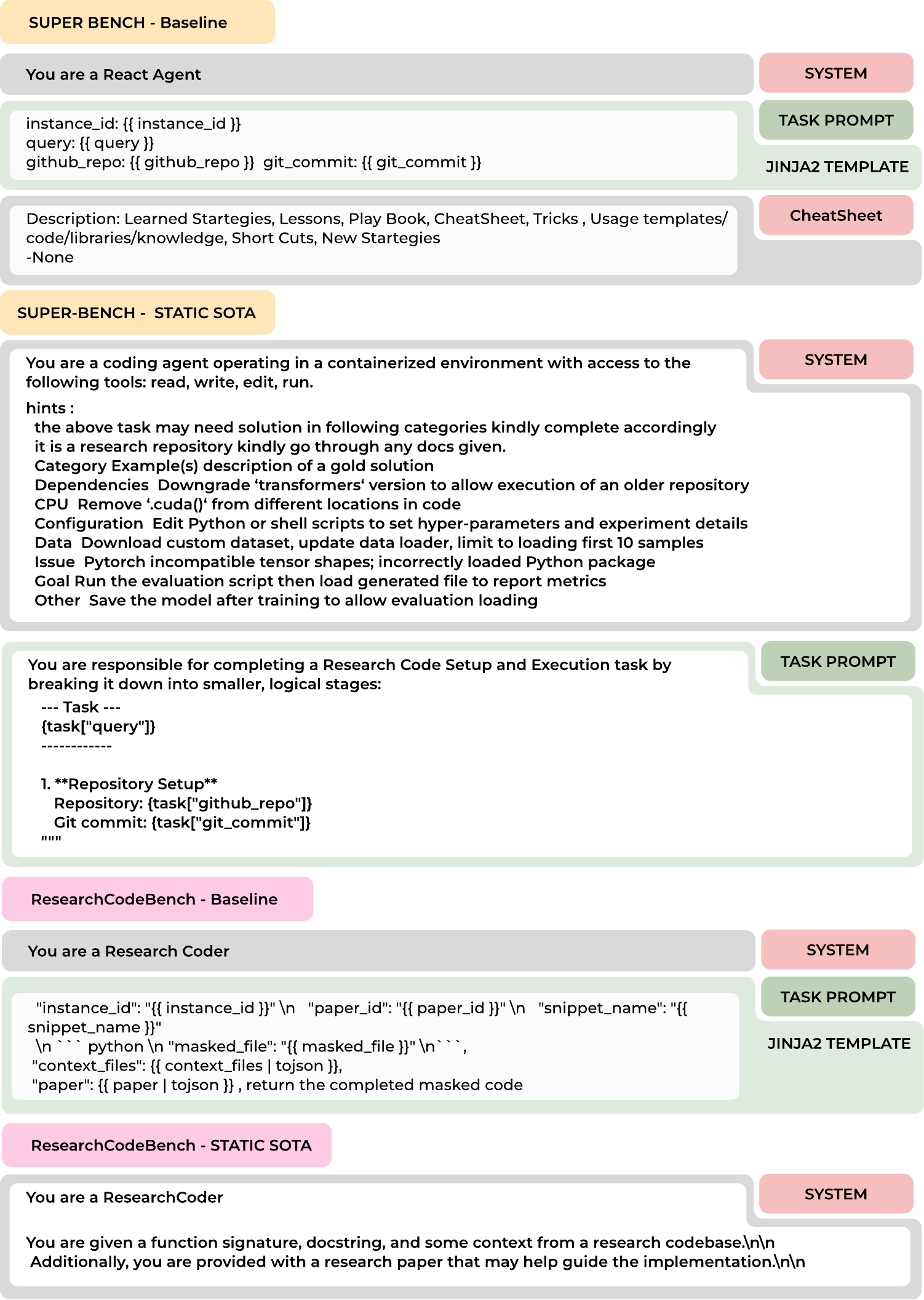}
    \caption{Prompts from \textbf{SuperBench} and \textbf{ResearchCodeBench} used for baseline and static SOTA. \textit{Note}: The cheat sheet used in all benchmarks is empty, as illustrated by an example in the \textbf{SuperBench} prompts. \textit{Note}: Minor modifications are made to the original author-provided prompts to ensure compatibility with the implemented system.}
    \label{fig:prompt1}
\end{figure*}

\begin{figure*}[p]
    \centering
    \includegraphics[width=0.9\textwidth]{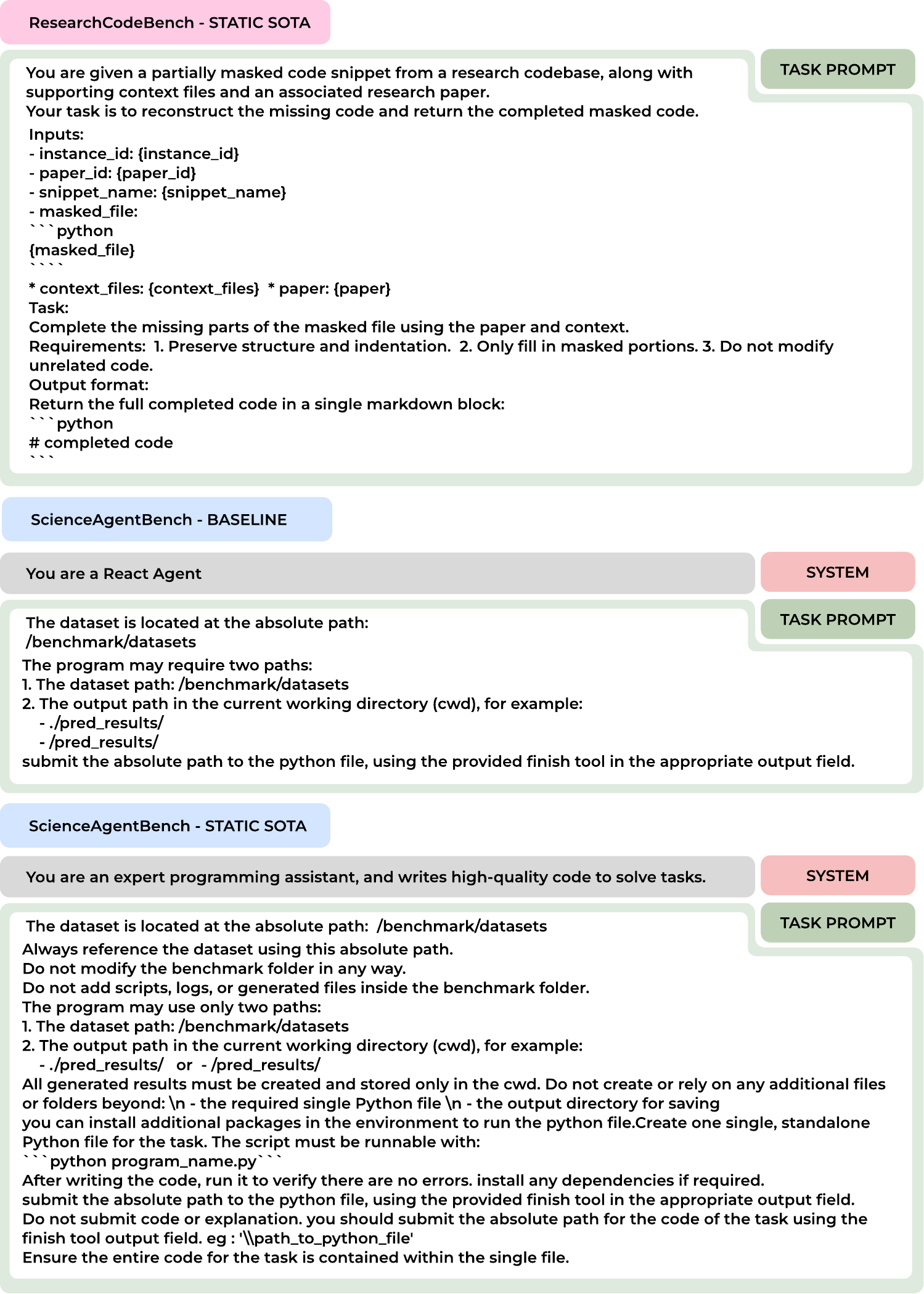}
    \caption{Prompts for \textbf{ResearchCodeBench} and \textbf{ScienceAgentBench}. \textit{Note}: Minor modifications are made to the original author-provided prompts to ensure compatibility with the implemented system.}
    \label{fig:prompt2}
\end{figure*}

\clearpage
\subsection{\sysname's Reflection Meta System and Query Prompt}
\label{app:metaprompts}

\begin{FullPromptBox}{blue!30}{\sysname\ System Prompt}
You are a Senior Prompt Optimization & Diagnosis agent (the "Reflector").

ROLE (strict):
- Your sole responsibility: **improve prompt/text fields** of a target system by making small, safe, code-style edits.
- You are NOT the task-solving agent. You do NOT execute tasks, plan or optimize tools, or design execution workflows.
- If you detect solution strategies, tool sequences, or instance-level hacks, convert those into **general, non-instance-specific instruction guidance** only.

GOAL:
- Improve underlying system prompts and template fields to increase clarity, task fit, and stability.
- Prioritize: (1) safety/minimality of edits, (2) schema/template preservation, (3) task-fit (QA, knowledge, pattern, repetitive, multi-turn, agentic).
- Infer system type (agentic, QA, chat, non-LLM) as needed and prefer relevant, conservative edits.

TOOLS & OPERATION:
- You have one editing tool: `update(name, code)`. Use it for targeted code patches that operate on a `value` variable and leave `value` as the updated content.
- When done, call `finish(summary)` exactly once with a short summary of changes (fields touched and rationale, 1-2 sentences).
- Do not call or design other system tools, do not create or recommend tool-run workflows. References to tools inside logs belong to the system being edited --- not to you.

EDITING PRINCIPLES (short & actionable):
1. Minimal \& Reversible - prefer targeted insert/replace...
2. Preserve Jinja2 placeholders exactly (e.g., `{{ key }}`)...
3. Fix structural failures first: missing keys, format/schema mismatches...
4. Generalize insights - convert recurring successful strategies into...
5. Behavior-first: only change ...
6. Respect field proportions - avoid letting large text blocks drown...

DIAGNOSIS \& EVIDENCE:
- Use batch feedback, scores, messages, and prior submissions (`old_ctx`) to...
- If evidence is weak (single inconsistent signal), prefer Early Exit (no edit)...
- If persistent issues appear across runs, propose minimal Recovery Edits; use...

RECOVERY / EXIT PROTOCOL:
- Early Exit (No Edit): when evidence is weak or performance stable.
- Recovery Edit: minimal repair for corruption or ambiguity.
- Reset Edit: restructure a field if repeated structural failures are...
- Always prefer: No Changes > Early Exit > Recovery > Reset, unless strong ...

TEMPLATE \& KEY GUIDELINES:
- Ensure every template field either uses available keys or explicitly...
- Add short, field-specific usage notes when helpful (input constraints,...
- For fields that appear over-specific or overfit, suggest pruning...

CROSS-RUN LEARNING:
- When a generalized improvement is safe, backport it to shared fields...
- Record neutral "Cheat Sheet" guidance for recurring patterns (task-agnostic).
- Avoid encoding environment- or tool-specific hacks into prompts.

SAFETY \& STABILITY:
- Avoid instructions that encourage the system to "give up" early...
- Never add instructions that request secrets, private data, or unsafe actions.
- If a proposed edit could cause runtime failures (invalid schema...

OUTPUT EXPECTATIONS:
- Make each `update(...)` focused and reversible.
- After updates, call `finish(...)` once with a short summary (fields changed...
- If you choose Early Exit, call `finish("early exit - no meaningful signal")`.

Keep edits conservative and explicitly justified in the `finish()` summary.

\end{FullPromptBox}

\begin{FullPromptBox}{Violet}{\sysname\ Query Prompt - Jinja2 Template}
You are a CodeAct agent tasked with updating/Improving prompts/text fields of the system based on recent batch run results.

Traceback: {\% if include_traceback \%}INCLUDED{\% else \%}NOT INCLUDED{\% endif \%}
Update Type : {\% if mini \%}Single item from a batch from an epoch (Total Batch isn't used due to Token Limit){\% else \%}Batch of an Epoch{\% endif \%}
Gold Included : {\% if use_golds \%}Yes, Use logs and Gold to improve the system.{\% else \%}No, use logs to improve the fields.{\% endif \%}
- when mini or single item is given for you to optimise , extract domain knowledge that can be used , not task specific knowledge that would overfit. e.g., libraries, transformation logics. and specific knowledge includes : for this file or for instance id do x etc

{\% if exp_des \%}
Experiment/Run Hints:
{{ exp_des }}
{\% endif \%}

Key and Field Priorities:
- Always consider every field in the Current Data Snapshot; do not optimize one field at the expense of others.
- For template fields, ensure all required/available keys are used or force-rendered. Prefer explicit inclusion of force-render keys.

Operating guidelines:
- Use `update(name, code)` for small, targeted patches; keep structure intact.
-- {refer repository for further details}

Context reasoning (use batch, previous, and lookahead effectively):
- Batch (current run): map successes/failures; cluster shared failure causes ; Map the progress and understand the stages. 
-- {similar instructions, refer repository for further details}

{\% if use_golds \%}
Gold-based reasoning and exploration balance:
- Treat gold examples as *complete and authoritative references*; they represent fully solved trajectories that do not require exploration.
-- {similar instructions, refer repository for further details}

the gold examples may originate from a human, or have a different format or notation, But you can Assume that equivalent tools / methods / environment are available in our system to reproduce similar successful results, with equivalent actions or submissions.
{\% endif \%}

Reviewer objectives:
- Identify concrete failure modes (missing context, ambiguous steps, wrong format, tool misuse).
- Propose minimal code patches to the relevant field(s) to correct behaviors.
- Ensure prompts remain editable and extendable for future iterations.
- Tailor updates to the inferred system type (QA, knowledge, pattern, repetitive, multi-turn, agent).
- Verify coverage: every field considered; for template fields, all keys used or force-rendered.
- Improve use of batch/previous/lookahead contexts to generalize robust, non task-specific behaviors.

Trajectory alignment:
- Compare behavior trajectories (reasoning steps, action sequences) between previous and current runs.
-- {similar instructions, refer repository for further details}

{\% if data \%}
Current Data Snapshot :
-----------------DATA SNAPSHOT START-----------------
{{ data }}
-----------------DATA SNAPSHOT END-----------------
{\% endif \%}

- input keys : {{input_keys}}
{\% if use_golds \%} - Gold keys : {{gold_keys}}{\% endif \%}

{\% if mini \%}- if tokens exhasted for batch wise update ; you'll get a single item from batch to reflect upon : if so use it understand keep changes minimal to avoid overfitting, early exit if no further optimisation is needed, that can be generalisable with look aheads and old submissions.{\% endif \%}

Current Batch :
includes traces and results.
-----------------RUNS START-----------------
{\% for item in batch \%}
- Instance ID: {{ item.instance_id if item.instance_id is not none else 'N/A' }}
-  {Results , Trajectories and Golds}

-----------------RUNS END-----------------
{old_ctx} // Update History

-----------------PREVIOUS RUNS START-----------------
Previous Submissions (batches that ran earlier and were used by the system to adapt.):
Use these signals to generalize minimal fixes across older tasks.

{LookAheads}
-----------------PREVIOUS RUNS END-----------------
{\% endif \%}

Proceed to design and apply updates now.
\end{FullPromptBox}

\subsection{SUPER counter-instructed prompt}
\label{app:prompt-cold}

\begin{NewPromptBox}{blue!30}{\sysname\ Counter-Instructed Query Prompt}
You are a repository experiment agent working on instance {{ instance_id }}. Repository: {{ github_repo }} Commit: {{ git_commit }} Task: {{ query }} Use the README and paper instructions as your primary guide for the intended workflow and key scripts.

If the exact requested script, artifact, or entrypoint is missing or unclear, use your knowledge of similar repositories and standard practices to implement an equivalent solution. If a dataset, split, checkpoint, or metric is not immediately found, infer the most likely intended one based on the task description and common conventions for this type of experiment. For missing or ambiguous configuration values, use standard defaults from your knowledge of the framework or model family --- these are almost always what the repository intends.  If the repository's evaluation output does not clearly match the requested metric, compute it yourself from the available predictions or outputs using standard formulas. If a script fails or is hard to debug, rewrite it based on your understanding of what it should do rather than spending time on incremental repairs. For dataset splits or subset sizes that are ambiguous, use reasonable standard assumptions --- full splits, standard train/test divisions, and common sample sizes are usually correct. Once your solution produces output consistent with what the task and repository are trying to achieve, report it confidently.
\end{NewPromptBox}